\documentclass[12pt]{article}

\usepackage{hyperref}

\usepackage[utf8]{inputenc}
\usepackage{amsmath}
\usepackage{amsthm}
\usepackage{amsfonts}
\usepackage{mathtools}
\usepackage{amssymb}
\usepackage{comment}
\usepackage{dsfont}
\usepackage[ruled,vlined]{algorithm2e}
\usepackage{pdfpages}

\usepackage{accents}

\usepackage{caption}
\usepackage{dsfont}
\usepackage{environ,wrapfig}
\usepackage{wasysym}
\usepackage{enumitem}
\usepackage{adjustbox}
\usepackage{ragged2e}
\usepackage{longtable}
\usepackage{changepage}
\usepackage{setspace}
\usepackage{hhline}
\usepackage{multicol}
\usepackage{tabto}
\usepackage{float}
\usepackage{multirow}
\usepackage{makecell}
\usepackage{fancyhdr}
\usepackage{mathdots}
\usepackage{mathtools}

\usepackage{amsthm}

\usepackage{times}

\usepackage{graphicx}
\usepackage{sidecap}

\mathtoolsset{showonlyrefs}

\usepackage{caption}
\usepackage{subcaption}

\newenvironment{sciabstract}{%
\begin{quote} \bf}
{\end{quote}}

\usepackage{mathtools}

\mathtoolsset{showonlyrefs}

\NewEnviron{wrappingcaptionfigure}[1]{
\stepcounter{figure}
\begin{wrapfigure}{r}{0.6\textwidth}
#1
\end{wrapfigure}
\noindent{}Figure \thefigure:
\BODY}

\newtheorem*{Th*}{Theorem}

\title{\Large\bf When social influence promotes the wisdom of crowds}

\author
{Abdullah Almaatouq$^{a,\ast\dagger}$, M. Amin Rahimian$^{b\dagger}$, Jason W. Burton$^{c}$,\\ and Abdulla Alhajri$^d$\\
\normalsize{$^a$~Sloan School of Management, Massachusetts Institute of Technology \vspace{-3mm}}\\
\normalsize{$^b$~Department of Industrial Engineering, University of Pittsburgh \vspace{-3mm}}\\
\normalsize{$^c$~Department of Psychological Sciences, Birkbeck, University of London \vspace{-3mm}}\\
\normalsize{$^d$~Department of Physics, University of Oxford \vspace{-3mm}}\\
\normalsize{$^\dagger$~A. Almaatouq and M.A.R contributed equally to this work.\vspace{-3mm}}\\
\normalsize{$^\ast$~To whom correspondence may be addressed; email: amaatouq@mit.edu.}
}

\date{}

\begin{document} 

\baselineskip18pt

\maketitle 

\begin{sciabstract}
Whether, and under what conditions, groups exhibit ``crowd wisdom'' has been a major focus of research across the social and computational sciences. Much of this work has focused on the role of social influence in promoting the wisdom of the crowd versus leading the crowd astray, resulting in conflicting conclusions about how the social network structure determines the impact of social influence. Here, we demonstrate that it is not enough to consider the network structure in isolation. Using theoretical analysis, numerical simulation, and reanalysis of four experimental datasets (totaling $2,885$ human subjects), we find that the wisdom of crowds critically depends on the interaction between (i) the centralization of the social influence network and (ii) the distribution of the initial, individual estimates. By adopting a framework that integrates both the structure of the social influence and the distribution of the initial estimates, we bring previously conflicting results under one theoretical framework and clarify the effects of social influence on the wisdom of crowds.
\end{sciabstract}

\newpage

\section*{1 \;\; Introduction}

 In its classical definition, the concept of ``the wisdom of crowds'' refers to the idea that the aggregate estimate of a group of individuals can be superior to that of individual, credentialed experts~\cite{galton1907vox,surowiecki2005wisdom}. Recent applications of this concept include technological, political, and economic forecasting~\cite{wolfers2004prediction}, crowdsourcing~\cite{vermeule2009many}, and public policy design~\cite{morgan2014use}. Conventional statistical accounts of the wisdom of crowds rely on the following two assumptions: (i) the individual errors are uncorrelated or negatively correlated~\cite{davis2014crowd}, and (ii) the individuals are unbiased, i.e., correct in mean expectations~\cite{surowiecki2005wisdom}.

However, social influence processes, in which people exchange information about their estimates, can cause them to revise their judgment in estimation tasks~\cite{muchnik2013social,lorenz2011social,becker2017network,golub2010naive}. Therefore, aggregating the revised (post-influence) estimates is not the same as aggregating the initial (pre-influence) estimates. Prior research yields conflicting findings on the effects of social influence on the wisdom of crowds. For instance, despite the evidence that social influence can significantly benefit group and individual estimates~\cite{becker2017network,bahrami2010optimally, gurccay2015power, becker2019wisdom,almaatouq2020adaptive}, social influence has also been found to induce systematic bias, herding, and groupthink~\cite{muchnik2013social, lorenz2011social}.

In response to these inconsistencies, notable reconciliation efforts have focused on investigating how social network theories interact with the process of collective belief formation. The results of these efforts, including seminal theoretical works~\cite{golub2010naive,demarzo2003persuasion} and laboratory experiments~\cite{becker2017network}, have established that the wisdom of crowds is preserved only if the influence of the most influential individual vanishes, i.e., becomes negligible, as the group size grows~\cite{golub2010naive}. This condition is satisfied in \textit{decentralized influence structures}, i.e., structures where everyone has an equal voice, as opposed to \textit{centralized structures} where one or more individuals have disproportionate influence. Intuitively, the wisdom of crowds benefits from larger group sizes, but centralized influence diminishes this benefit by reducing the collective estimate to the ``wisdom of the few."

While these results appear to broadly suggest the superiority of decentralized influence, their conclusions rest on the premise that the distribution of the initial estimates is centered around the truth. In such situations, there are no opportunities for the crowd to improve with social influence~\cite{golub2010naive}. However, empirical distributions of numerical estimates tend to be right-skewed with excess kurtosis, where most estimates are low, with a minority falling on a fat right tail~\cite{lorenz2011social,jayles2017social, kao2018counteracting}. The skewness of the distribution could emerge due to systematic bias (a tendency to over- or underestimate the actual value~\cite{jayles2017social,indow1977scaling,simmons2011intuitive}) or dispersion (the spread of the estimates) in the population. Therefore, it is when the crowd is not initially centered around the truth, as observed in many empirical settings, that centralized influence could present an opportunity to promote crowd wisdom.

In this study, we ask \textit{when centralized influence structures improve or hinder the wisdom of crowds in estimation tasks}. Our results demonstrate that the effect of social influence varies systematically with the distribution of the initial estimates, and, therefore, it is more heterogeneous than previously suggested. Specifically, we analyze ---  theoretically, numerically, and empirically --- the effect of the distribution of initial estimates on the suitability of a crowd to benefit from influence centralization. 

\section*{2 \;\; Theoretical Model}

\begin{figure}[tbh]
\includegraphics[width=1\textwidth]{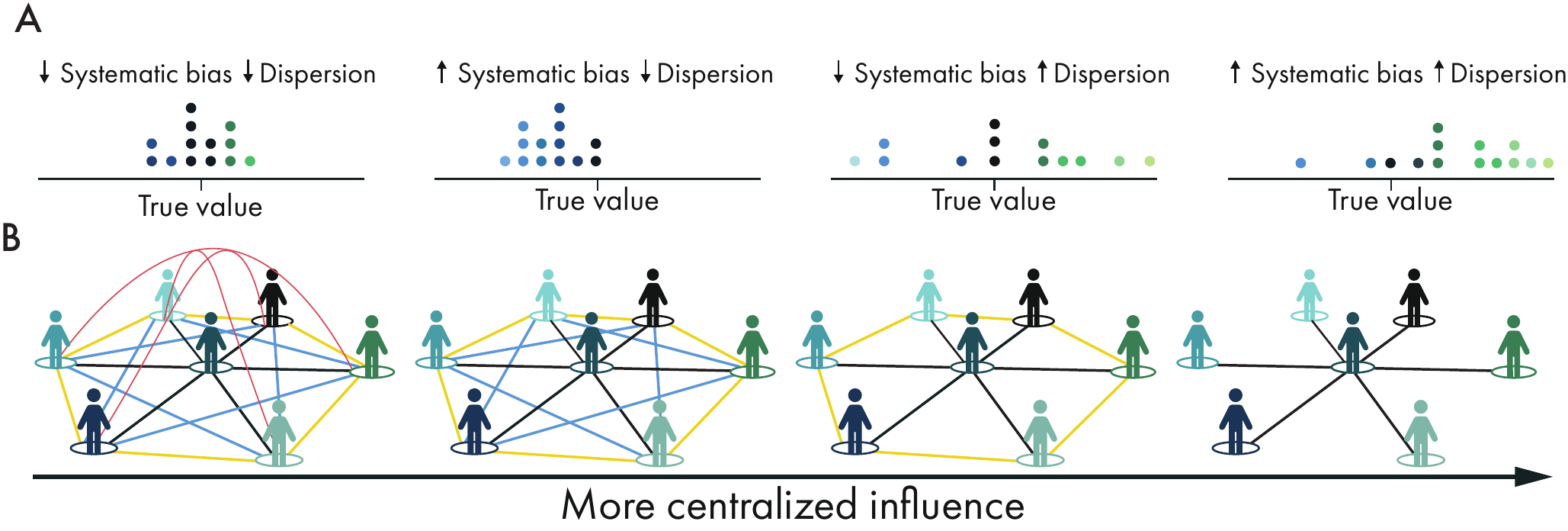}
\caption{\textbf{This schematic illustrates our framework for analyzing the role of estimation context in determining how social influence shapes the wisdom of crowds.} Panel \textbf{A} illustrates four distributions of the initial estimates. Panel \textbf{B} provides examples of different influence network structures arranged in the order of increasing centralization --- from a fully decentralized structure, where everyone has an equal voice, to a highly centralized structure, where there is one highly influential individual.}
\label{fig:problem_schematic}
\end{figure}

To illustrate this, we consider a group of $n$ agents tasked to estimate or forecast, with maximal accuracy, some unknown positive quantity such as the unemployment rate in the next quarter, life expectancy of an ill patient, amount of calories in a meal, prevalence of global influenza infections in two weeks, or number of jellybeans in a jar. To model the population of the agents performing a particular estimation task, we endow each agent with a biased and noisy signal about the truth that constitutes her initial estimate. Let the group of $n$ agents be indexed by $i = 1, \ldots, n$, and denote their initial estimates by $\mathbf{a}_{i,0}$. The initial estimates are independent and identically distributed  and their common distribution, $\mathcal{F}_{\mu,\sigma}^{\theta}$, is parametrized by the unknown truth, $\theta$, the systematic bias, $\mu$, and the dispersion, $\sigma$.  The location parameter ($\mu$) indicates the center of the distribution that biases the estimates with respect to the truth, and the shape parameter ($\sigma$) determining the variation and tail shape. The skewness of the distribution can emerge due to several possibly interdependent factors: disproportionate exposure to a skewed sample of task instance~\cite{dehaene2008log, resnick2017dealing}, the  tendency to over-attend to the information that supports one's hypotheses~\cite{oswald2004confirmation}, or the level of demonstrability of the task at hand~\cite{jayles2020impact}. In general, the initial estimates can be viewed as intrinsic properties of the \textit{estimation context}: a population of agents performing a particular estimation task instance. Different populations of agents, e.g., experts vs. novices, might have different biases and dispersion for the same task instance. Conversely, the same population can vary in terms of their bias and dispersion across different task instances. For brevity and to abstract the agents and the estimation task, we refer to the distribution of the initial estimates as the \textit{estimation context}. Figure~\ref{fig:problem_schematic}.A shows four estimation contexts with varying levels of bias and dispersion.

Agents frequently have access to the opinions or estimates of other agents. We define the collective estimate of the $n$ agents as the average of their revised (post-influence) estimates and denote it by ${\mathbf{a}}^{n}$. In many common models of social influence~\cite{golub2010naive,demarzo2003persuasion,DeGrootModel}, as well as in other aggregation mechanisms~\cite{davis2014crowd,anderson2014functional, prelec2017solution}, the collective estimate of the group of agents can be expressed as a convex combination (weighted average) of the initial estimates: ${\mathbf{a}}^{n}(\bar{w}) = \sum_{i=1}^{n}w_i\mathbf{a}_{i,0}$, where $w_1,\ldots,w_n$ are positive real weights summing to one. These weights represent the influence of individual agents on shaping the collective estimate. Without loss of generality, we assume that the agents are ordered in the decreasing order of their influence, i.e., $w_1 \geq w_2 \geq \ldots \geq w_n$. This definition of collective estimation contains the simple average of the initial estimates as a special case, i.e., the typical ``wisdom of crowds''. 

We introduce an influence \textit{centralization} parameter, $\omega$, to interpolate between a collective estimate produced by a fully decentralized influence setup where every agent has an equal voice  (i.e., $\omega = 0$ and $w_1 = w_2 = \ldots = w_n = 1/n$), $\omega=0$, and a dictatorial setup with a single influential agent (i.e., $w_1 =\omega = 1$ and $w_2 = \ldots = w_n = 0$), $\omega=1$. In order to investigate the role of network centralization, $ 0 \leq \omega \leq 1$, we consider a class of influence structures indexed by $\omega$ such that (see SI section S1.1 for more details),
\begin{align}
    {\mathbf{a}}^{n}(\omega) = \omega\mathbf{a}_{1,0} + (1-\omega)\frac{1}{n}\sum_{i=1}^{n}\mathbf{a}_{i,0}. \label{eq:collectiveestimateMM}
\end{align} Our definition of $\omega$ coincides with Freeman's centralization~\cite{freeman1978centrality} for a class of network typologies that encompass cases of practical and empirical interest, such as fully connected networks, star networks, empty graphs (isolated individuals), and circular lattices, among others. Figure~\ref{fig:problem_schematic}.B shows four influence network structures in this class (see SI section S1.1 for calculation of $\omega$ in different networks). It is important to note that these networks are \textit{influence networks} (tie between two people is represented as a weighted value between zero and one) and not \textit{communication networks} (i.e., binary networks that define who communicates with whom).

We measure the collective performance of the agents in terms of the proximity of the collective estimate (${\mathbf{a}}^{n}$) to the truth ($\theta$). Given the estimation context (distribution of the initial estimates), our outcome of interest is whether the collective estimate produced by a centralized influence structure outperforms a decentralized baseline. We compute the probability of this outcome for a given estimation context and denote it by $\Omega_n$. Notably, $\Omega_n$ captures a critical \textit{feature} of the estimation context, namely, \textit{its suitability to benefit from centralization}. For instance, when $\Omega_n < 1/2$, the initial estimates are better suited for decentralized influence structures; conversely, when $\Omega_n > 1/2$, they are better suited for centralized influence structures.

\section*{3 \;\; Results}

 \textbf{Analytical results.} Our theoretical analysis of $\Omega_n$ verifies that for heavy-tailed or right-skewed distributions, the performance of the collective estimate in a centralized structure where a single agent has a non-vanishing influence (her contribution to the collective estimate does not go to zero as $n\to\infty$) is superior to that of the decentralized baseline. In particular, for heavy-tailed distributions (e.g., Pareto, log-normal, and log-Laplace), we identify phase transition behaviors, whereby the lower bound's limiting value transitions from $0$ to $1$ or $1/2$, as the shape parameter, $\sigma$, crosses a critical value (see SI section S2.1).  Intuitively, this is due to the fact that the sample mean of a heavy-tailed distribution is dominated by its excess tail risk (the egregious errors of a few individuals) in decentralized networks. On the other hand, with weighted averages as in centralized structures, we can guarantee that some random individuals exert enough influence to prevent the group aggregate from being swayed too far by the egregious errors of the few.
Notably, in this model, centralized structures violate the vanishing influence condition for the wisdom of crowds, cf. \cite{golub2010naive} and SI section S2.1.5. This underscores the importance of the distributional assumptions, which are context dependent, when studying the effect of social influence on the wisdom of crowds.

\begin{figure}[H]
\vspace*{-5pt}
\centering
\includegraphics[width=0.65\textwidth]{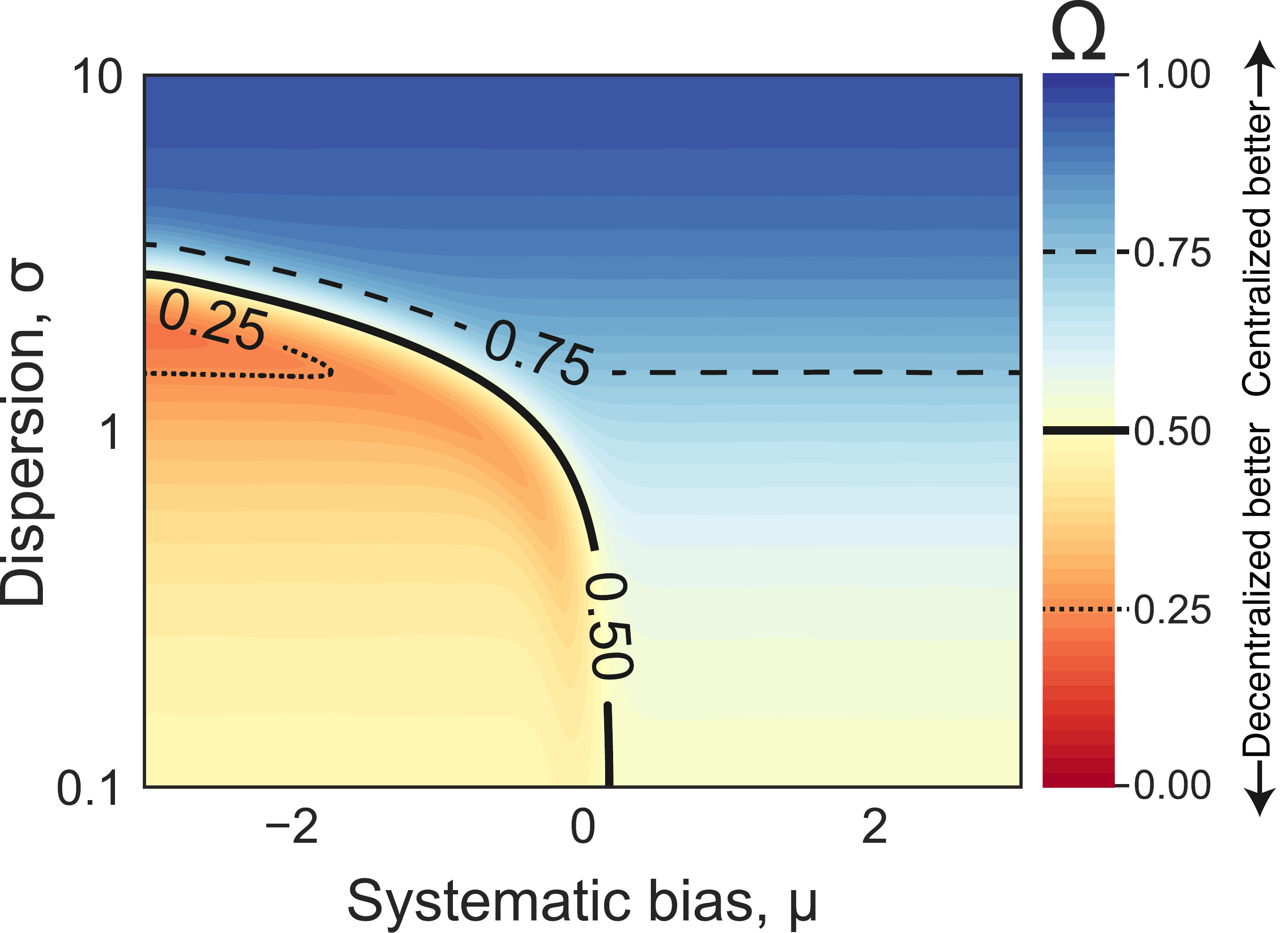}
\caption{\textbf{The link between the distribution of the initial estimates and the probability that collective estimation on a centralized structure outperforms a decentralized structure ($\Omega_n$)} Our outcome of interest, $\Omega_n$, is the likelihood that a weighted average falls closer to the truth than an unweighted average. Hence, when $\Omega_n < 1/2$, the estimation context is better suited for decentralized (unweighted) influence structures; conversely, when $\Omega_n > 1/2$, it is better suited for centralized (weighted) influence structures. In this figure, the initial estimates are sampled from a log-normal distribution while varying location and shape paraneters ($\mu$ and $\sigma$). The number of agents and influence-centralization level are fixed at $n=50$  and $\omega = 1/3$, respectively. See SI Figures S1 and S3 for other distributions and parameter choices.}
\label{fig:phase_diagram}
\end{figure}

In Figure~\ref{fig:phase_diagram}, we illustrate the behavior of $\Omega_{n}$ for a log-normal distribution of initial estimates, as reported in several empirical studies~\cite{lorenz2011social,becker2017network,kao2018counteracting}. In this case, $\Omega_{n}$ predicts that centralized influence structures improve collective estimates over decentralized influence structures if the distribution of the initial estimates is characterized by overestimation bias or large dispersion (see SI
section S2.2.1 and Figure S2 for the effect of the systematic bias). However, this relationship is reversed when the distribution is characterized by low dispersion and underestimation bias (see SI section S2.2 and Figure S1 for simulation details and other distributional classes).

\begin{figure}[H]
\vspace{-50pt}
\begin{center}
\includegraphics[width=0.7\textwidth]{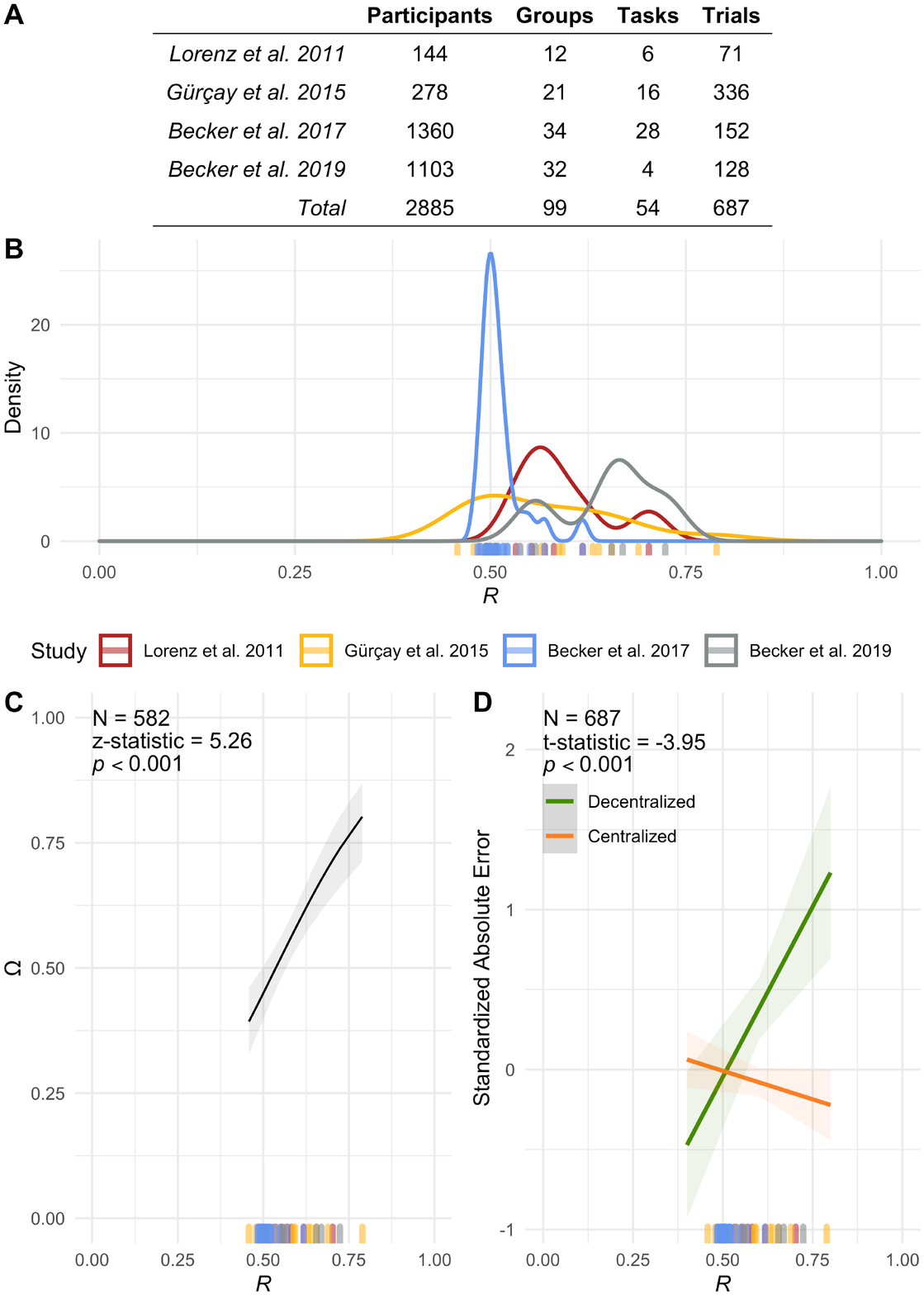}
    \caption{\textbf{Reanalysis of previously published experiments indicates that our proposed feature, $R$, has significant predictive powers for determining when the group performance improves as a result of social interactions.} Panel \textbf{A} shows the number of participants, groups, tasks, and trials in the reanalyzed experiments. Panel \textbf{B} displays the distribution of $R$ across these studies. Panel \textbf{C} shows that the probability of groups improving their performance after social interaction, $\Omega$, is substantially explained by $R$. Panel \textbf{D} shows the marginal effect of the interaction term between influence centralization and $R$: as $R$ increases, the group performance improves in the centralized influence conditions, and degrades in the decentralized influence conditions. The bands are the $95\%$ confidence intervals.}
\label{fig:empirical_data}
\end{center}
\end{figure}

\textbf{Reanalysis of four experimental datasets.} In order to empirically test the predictions of the aforementioned model, we use the data pertaining to positive numerical estimation tasks (i.e., tasks with negative estimates were omitted) from four published experiments~\cite{lorenz2011social, becker2017network, gurccay2015power, becker2019wisdom}. In these experiments, a total of 2,885 participants, organized into 99 independent groups, completed a total of 54 estimation tasks, generating a total of 15,562 individual estimations and 687 collective estimations (see Figure~\ref{fig:empirical_data}.A).  

All experiments followed a similar procedure that involved the following three steps: (1) the participants simultaneously and independently completed numeric estimation tasks on a range of topics, e.g., visual estimation, trivia questions, political facts, and economic forecasts; (2) within groups of varying sizes, the participants in the social interaction condition communicated information about their estimates with each other; and (3) the participants had one or more opportunities to revise their estimates. One trial consisted of a single group of participants answering a single task.

Each task induces a different distribution on the initial estimates that are observed empirically. For each task, we use the relative log likelihood of a fitted log-normal distribution versus a normal distribution (hereinafter denoted by $R$), as a measure of the heavy-tailedness of the distribution of initial estimates. In other words, this measure captures whether the initial estimates for a given task are better described by a thin-tailed distribution (i.e., normal distribution) or a heavy-tailed one (i.e., log-normal distribution). That is, $R=0$ indicates with 100\% certainty that the initial estimates are better fit by a normal distribution than a log-normal distribution, $R=1$ indicates with 100\% certainty that the initial estimates are better fit by a log-normal distribution than a normal distribution, and $R=0.5$ indicates that the initial estimates could equally be described as normally or log-normally distributed. Figure~\ref{fig:empirical_data}.B shows the distribution of the empirically derived $R$ in these studies, which confirms that the majority of the empirical distributions of the initial estimates are better described by a heavy-tailed distribution than a thin-tailed one.

We refer to the average estimate of the individuals in each group, \textit{before} and \textit{after} their interactions, as their collective \textit{initial} and \textit{revised} estimates, respectively. For each trial, we compare the absolute errors of the collective initial and revised estimates, and use the following two outcome metrics: (1) whether the collective revised estimate is more accurate than the collective initial estimate in groups with social interaction, and (2) the standardized (z-score) absolute error of the revised collective estimate for all groups (with or without social interaction). We use a logistic regression for the former, and a linear regression for the latter.

This empirical analysis relies on the following premise: the collective initial estimate corresponds to the most decentralized influence structure ($\omega=0$), and social interactions can only increase the influence centralization ($\omega>0$). For example, even in social interactions where everybody is equally connected in terms of the communication structure, some group members may become more influential than others, by virtue of being more talkative~\cite{dalkey1969experimental}, more persuasive, or more resistant to social influence~\cite{becker2017network,almaatouq2020adaptive}. The key insight is the fact that the collective initial estimate (pre-social interaction), eliminates the possibility of any variation in influence and therefore it is equivalent to the most decentralized network. In contrast, the collective revised estimate (post-social interaction) can be influenced disproportionately by domineering individuals, and therefore can be modeled as a centralized influence network. The same insight can be extended to modeling unstructured discussion as centralized influence and the Delphi method (and other mediated communication techniques) as a relatively decentralized networks. (The interested reader can find a more careful discussion in a follow-up to this paper by Becker et al. (2020)~\cite{becker2020network} showing  the application of this modeling insight and our results here to explain why unstructured discussion will sometimes outperform numeric communication and why the outcome is sometimes reversed).

Here, we begin by testing the main hypothesis predicted by our theory; namely, that the effect of social influence centralization on the performance of groups is moderated by our measure of the heavy-tailedness of the distribution of initial estimates, $R$. As shown in Figure~\ref{fig:empirical_data}.C, we find that the probability that a group improves after centralized social interaction --- denoted by $\Omega$ as the outcome variable of interest --- is substantially explained by $R$ (z-statistic $=5.26$; $p<0.001$).

In Figure~\ref{fig:empirical_data}.D, we find that the interaction between the centralization of the influence and the measure of the heavy-tailedness of the distribution of initial estimates, $R$, significantly affects the the absolute error of the revised collective estimate ($\beta$ $= -4.97$; t-statistic $=-3.95$; $p < 0.001$). Critically, the results of this empirical analysis show that variation in the heavy-tailedness of the distribution of initial estimates, $R$, can completely reverse the effects of social influence centralization: when $R<0.5$, the error of the revised collective estimate is lower in decentralized influence structures; whereas when $R>0.5$, the error of the revised collective estimate is lower in centralized structures (Figure \ref{fig:empirical_data}.D).

\section*{4 \;\; Discussion}

The primary contribution of this paper is to reconcile previous research about a question that is fundamental to understanding the performance of groups: how does social influence impact the accuracy of collective estimates? The critical implication of our results is that the attributes of the districution of the intial estimates(i.e., the estimation context) moderates the effect of influence centralization. Therefore, we find no support to the hypothesis that decentralized influence structures are preferred to centralized ones \textit{independently of the estimation context}.

Therefore, the effect of network structure on the collective estimation performance should be reconceptualized under a context-dependent framework, i.e., with respect to the population of individuals performing the particular task. There is no single influence structure that is better than others in all contexts. Such a context-dependent framework can unify previously conflicting findings on crowd wisdom under a single theoretical framework and explain the effects of the influence network structure on the quality of the collective estimates.

Admittedly, the estimation context is only one of several potential sources of inconsistency in previous studies. For instance, vagueness or ambiguity of some theoretical constructs, e.g., influence, can result in different studies of, seemingly, the same phenomenon, to measure different things.

Although the calculation of our proposed feature of the estimation context, $R$, does not require knowledge of the truth (estimand), it does require access to a group's set of initial estimates. However, we note that research on simple estimation tasks demonstrated that similar classes of estimation tasks tend to yield similar and reliably predictable distributions of initial estimates~\cite{kao2018counteracting}. Thus, prescriptively, for a group that may regularly need to complete the same class of estimation tasks routinely, it may be possible to calibrate group structure based on historical data and long term feedback. 

Furthermore, we acknowledge that our outcome of interest, $\Omega$, is concerned only with the probability of the following event: the collective estimate generated by the agents interacting in a centralized influence structure is closer to the truth than the collective estimate generated by the agents in a decentralized structure. This is not the same as comparing mean squared error or other expected loss functions for these collective estimates (see Figure S4 for examples of other loss functions).

Finally, we note that we only studied one class of tasks: numerical estimation with a non-negative, objective truth. Relevant research on other classes of tasks has similarly demonstrated that variation in context features, such as complexity~\cite{barkoczi2016social,shore2015facts,mason2012collaborative,lazer2007network}, fundamentally alter  collective problem solving outcomes. 

To conclude, our theoretical and empirical analysis has demonstrated that conclusions about the role of the social influence can be inconsistent unless the estimation context is explicitly accounted for. Many research extensions are warranted from this framework. For example, unlike what is assumed in most available work, including ours, the social networks we live in are not random, nor are they imposed by external forces. Rather, these social networks emerge under the influence of endogenous social processes and gradually evolve within a potentially non-stationary context. A truly context-dependent view on crowd wisdom should open connections with diverse research fields and help advance an interdisciplinary understanding of the design of social systems and their information outcomes.


\section*{Materials and Methods}

\noindent\textbf{Theoretical analysis of $\Omega$.} We measure the probability that the collective estimate produced by a centralized influence structure, ${\mathbf{a}}^{n}(\omega)$, $\omega > 0$, outperforms the decentralized baseline, ${\mathbf{a}}^{n}(0)$. We denote this probability by $\Omega_n(\omega,\mathcal{F}^{\theta}_{\mu,\sigma}) := \mathbb{P}^{\theta}_{\mu,\sigma}[|{\mathbf{a}}^{n}(\omega) - \theta| < |{\mathbf{a}}^{n}(0) - \theta|]$. To compute $\Omega_n$ in Figure~\ref{fig:phase_diagram}, we have fixed $n=50$, $\theta = 2$, and $\omega = 1/3$. Therefore, $\Omega$ is entirely determined by the distribution of the initial estimates ($\mu$ and $\sigma$). Figure S3 replicates our simulation for a range of $n$ and $\omega$ values. For distributions $\mathcal{F}_{\mu,\sigma}^{\theta}$, supported over positive reals, with cumulative function ${F}_{\mu,\sigma}^{\theta}$, we propose the following lower bound (proved in SI section S2.1):
\begin{align}
    \Omega_n(\omega,\mathcal{F}_{\mu,\sigma}^{\theta}) \geq \displaystyle\sup_{\beta>\theta/(1-\omega)} \left\{ F^{\theta}_{\mu,\sigma}(\beta)(1 - {F}^{\theta}_{\mu,\sigma}(n\beta)^{n-1})\right\}. \label{eq:OmegaMM}
\end{align} In SI section S2.1, we show how to limit the rate of tail decay for different classes of distributions, to produce a non-trivial (non-zero) lower bound as $n \to\infty$. For heavy-tailed distributions, such as Pareto, log-Laplace, and log-normal (see SI subsections S2.1.1 to S2.1.3), we identify phase transition behaviors, whereby the proposed lower bound's limiting value transitions from $0$ to $1$ or $1/2$, as $\sigma$ crosses a critical value. \\ 


\noindent\textbf{Statistical tests.} All statistics were two-tailed and based on mixed-effects models that included random effects to account for the nested structure of the data. In particular, the logistic regression  for Figure~\ref{fig:empirical_data}.C is:
\begin{align}
    y_{ij} = \frac{1}{1+\exp(\beta_0 + \beta_1 R_{j} + v_i + \epsilon_{ij})}, \label{eq:genlinregMM}
\end{align} where $y_{ij}$ is a binary indicator for whether or not the $i$-th group in the $j$-th estimation context improved the accuracy of its collective estimate after social interaction; $\beta_0$ is the fixed intercept for the regression model; $\beta_1$ is the fixed coefficient for the estimation context feature, $R$; $v_i$ is the random coefficient for the $i$-th group; and $\epsilon_{ij}$ is a Gaussian error term. The analysis was conducted on 678 observations (groups with social influence).  

The regression equation for Figure~\ref{fig:empirical_data}.D is: 
\begin{align}
   y_{ij} = \beta_0 + \beta_1 R_{j} + \beta_2 I_{i} + \beta_3 I_{i} R_{j} + v_i + \epsilon_{ij},  \label{eq:linregMM}  
\end{align} where $y_{ij}$ is the standardized (z-score) absolute error of the revised collective estimate for the $i$-th group in the $j$-th estimation context, $R_{j}$; $\beta_0$ is the fixed intercept for the regression model; $\beta_1$ is the fixed coefficient for the estimation context feature, $R$; $I_{i} \in \{0,1\}$ is an indicator variable of whether or not social interaction has occurred; $\beta_2$ is the fixed coefficient for the social influence centralization; $\beta_3$ is the fixed coefficient for the interaction term between the estimation context feature, $R$, and influence centralization (shown in Figure~\ref{fig:empirical_data}.D); $v_i$ is the random coefficient for the $i$-th group; and $\epsilon_{ij}$ is a Gaussian error term. The absolute error of the revised collective estimate has been standardized, i.e., z-scored, in order to compare errors across different tasks (the correct answer for different tasks can differ by orders of magnitude). The analysis was conducted on 687 observations; 582 groups with social influence (centralized), and 105 groups without social influence (decentralized).

Further details of the regression analysis are provided in SI section S3.1, Table S1. Robustness checks for the regression results are presented in Tables S3-S2.  \\

\noindent\textbf{Data and code availability.} at \href{https://github.com/amaatouq/task-dependence}{https://github.com/amaatouq/task-dependence}.

\newpage

{\baselineskip0pt

\bibliographystyle{apalike}
\bibliography{ref}
}

\newpage



\section*{Supplementary material}

# Supplementary Information

This Supplementary Information is organized in five sections. In section S1, we present network models of collective estimation that facilitate our study of the collective estimate, $\mathbf{a}^n(\omega)$, as a function of the centralization, $\omega$, and in relation to its network structure. In section S2, we give the proof of our main proposition (lower bound on $\Omega_n$) and relate it to the rate of tail decay for many common distributions. We study these relationships theoretically and through numerical analysis as well. In section S3, we develop empirical measures to analyze prior experiments in terms of the features of their estimation context. In section S4, we provide robustness checks. Supplementary references are listed in section S5.

## Contents

---

**Preliminary notation.** For the convenience of the reader, we collect here some notation that will be used throughout. We study collective estimation by a group of agents and use $n$ to denote the group size. For sequences of real numbers $f_n$ and $g_n$, indexed by integers $n$, we use the asymptotic notations $f_n \asymp g_n$ to signify that $\lim_{n\to\infty} f_n/g_n = 1$. We consider agents that are indexed by $i = 1, \ldots, n$. We use bold fonts to represent random variables. We use $\mathbf{a}_{1,0}, \ldots, \mathbf{a}_{n,0}$ to denote the individual estimates in the absence of any social interactions. The collective estimate is denoted by $\mathbf{a}^n$; it is determined in terms of the individual estimates, in a manner that involves a centralization parameter $\omega$: $\mathbf{a}^n(\omega) = \omega \mathbf{a}_{1,0} + (1-\omega)\frac{1}{n}\sum_{i=1}^{n} \mathbf{a}_{i,0}$. Vectors are denoted by a bar on top of their letters and we use superscript $^T$ to denote matrix transpose.

# S1 Network models of collective estimation

Let $\theta$ be an unknown state of the world. Consider $n$ agents indexed by $i = 1, \ldots, n$, each endowed with a biased and noisy signal about $\theta$. The signals are independent and identically distributed across the agents and constitute their initial estimates of the unknown $\theta$:

$$\mathbf{a}_{i,0} \sim \mathcal{F}^{\theta}_{\mu,\sigma}. \tag{S.1}$$

The the distribution of the initial estimates in (S.1), $\mathcal{F}^{\theta}_{\mu,\sigma}$, is parametrized by $\theta$, $\mu$, and $\sigma$. We think of $\mu$ as a location parameter (the center of the distribution) that biases the individual estimates against $\theta$. This captures the level of systematic bias in the population. We think of $\sigma$ as a variance-proxy/shape parameter that determines the variation and tail-fatness of $\mathcal{F}^{\theta}_{\mu,\sigma}$. In other words, $\sigma$ can be interpreted as the amount of prior information a group has about the quantity and represents the level of *demonstrability* of the estimation task.

The agents interact in a group. Their group interactions can be modeled in a variety of ways leading to a group aggregate $\mathbf{a}^n(\bar{w})$ that is a convex combination of the initial estimates:

$$\mathbf{a}^n(\bar{w}) := \bar{w}^T \bar{\mathbf{a}}_0 = \sum_{i=1}^{n} w_i \mathbf{a}_{i,0}, \tag{S.2}$$

where $\bar{w}$ is an entry-wise non-negative vector satisfying $\bar{w}^T \mathbb{1} = 1$.

In general, different agents' initial estimates will receive different weights in the collective estimate. A common method of modeling group interactions is through DeGroot-style iterated averaging, which has a long history in mathematical sociology and social psychology [1]. The origins of iterated averaging models can be traced to French's seminal work on "A Formal Theory of Social Power" [2], followed up by Harary's investigation of the mathematical properties of the averaging model, including the consensus criteria, and its relations to Markov chain theory [3]. This model was further popularized by DeGroot's seminal work [4] on linear opinion pools and belief exchange dynamics. In a typical iterated averaging setup, an agent's estimate at time $t$ is given by a weighted average of her neighboring estimates at time $t - 1$:

$$\mathbf{a}_{i,t} = \sum_{j=1}^{n} W_{ij} \mathbf{a}_{j,t-1},$$

In matrix notation, we have:

$$\bar{\mathbf{a}}_t = W\bar{\mathbf{a}}_{t-1} = W^t \bar{\mathbf{a}}_0, \tag{S.3}$$

where $\bar{\mathbf{a}}_t = (\mathbf{a}_{1,t}, \mathbf{a}_{2,t}, \ldots, \mathbf{a}_{n,t})^T$ and $W = [W_{ij}]$ is the matrix of weights. We refer to matrix $W$ as the social influence matrix. For a strongly connected social network with $W_{ii} > 0$ for all $i$, the Perron-Frobenius theory [5, Theorems 1.5 and 1.7] implies that $W$ has a simple positive real eigenvalue equal to 1. Moreover, the left and right eigenspaces associated with the unit eigenvalue are both one-dimensional with the corresponding eigenvectors $\bar{w} = (w_1, \ldots, w_n)^T$ and $\mathbb{1} = (1, \ldots, 1)^T$. The magnitude of any other eigenvalue of $W$ is strictly less than 1; therefore, we have:

$$\lim_{t \to \infty} \bar{\mathbf{a}}_t = \lim_{t \to \infty} W^t \bar{\mathbf{a}}_0 = \mathbb{1}\bar{w}^T \bar{\mathbf{a}}_0 = \mathbb{1}\mathbf{a}^n(\bar{w}), \tag{S.4}$$

which implies a consensus on the collective estimate (S.2). In several experimental settings [6, 7, 8, 9], human participants get to revise their numerical estimates a few times only, and the collective estimate is then calculated by averaging the revised estimates. Let us denote the number of communication rounds in such a scenario by $\tau$. Using (S.3) to model the revision of the numerical estimates, we again arrive at a model that gives the collective estimate as a convex combination of the initial estimates, $\mathbf{a}^n(\bar{w}_\tau) = \bar{w}_\tau^T \bar{\mathbf{a}}_0$, where the transposed vector of weights, $\bar{w}_\tau^T$, is given by $\bar{w}_\tau^T = (1/n)\mathbb{1}^T W^\tau$.

## S1.1  $\omega$: Parameterizing a class of networks by their centralization

Motivated by our interest in comparing the collective estimation performance of centralized and decentralized networks, we focus our attention on a class of social influence network structures for which the collective estimate can be written as follows:

$$\mathbf{a}^n(\omega) = \omega \mathbf{a}_{1,0} + (1-\omega)\frac{1}{n}\sum_{i=1}^{n} \mathbf{a}_{i,0}, \tag{S.5}$$

where $\omega$ is a measure of influence centralization, with $\omega = 1$ representing a fully centralized social influence structure ($w_1 = 1$, and $w_2 = \ldots = w_n = 0$) and $\omega = 0$ corresponding to a fully decentralized social influence structure ($w_1 = \ldots = w_n = 1/n$). Indeed, with $\bar{w}$ as the centrality vector, the parameter $\omega$ corresponds to the Freeman centralization of the underlying

network with social influence matrix $W$. By varying $\omega \in [0, 1]$, we can interpolate between the two extremes: full centralization, $\omega = 1$, and complete decentralization, $\omega = 0$. This class, although not encompassing, is ideal for addressing the central question of interest in this work. The networks in this class are such that one agent, $i = 1$, is distinguished with a higher influence and all others, $i > 1$, have an equal but lower influence. Networks in this class include cases of practical and empirical interest, such as star networks and circular lattices. All networks in Figure 1.B, in the main text, belong to this class.

Here, we consider the special cases of star and cycle networks with bidirectional edges, and equal weights on all edges that are incoming to the same node. Their respective social influence matrices and associated left eigenvectors are given by:

$$W_\star = \begin{bmatrix} 1/n & 1/n & 1/n & \cdots & 1/n \\ 1/2 & 1/2 & 0 & \cdots & 0 \\ 1/2 & 0 & 1/2 & 0 \cdots & 0 \\ \vdots & \vdots & \ddots & \vdots & 0 \\ 1/2 & 0 & \cdots & 0 & 1/2 \end{bmatrix}, \quad W_o = \begin{bmatrix} 1/3 & 1/3 & 0 & \cdots 0 & 1/3 \\ 1/3 & 1/3 & 1/3 & 0 \ldots & 0 \\ 0 & \ddots & \ddots & & 0 \\ \vdots & & & & \\ 0 & \ddots 0 & 1/3 & 1/3 & 1/3 \\ 1/3 & \ldots & 0 & 1/3 & 1/3 \end{bmatrix},$$

$$\bar{w}_\star = (n/(3n-2), 2/(3n-2), \ldots, 2/(3n-2))^T, \quad \bar{w}_o = (1/n, 1/n, \ldots, 1/n)^T.$$

Subsequently, the network centralization parameter $\omega$ in (S.5) for the star and cycle networks are given by $\omega_\star = (n-2)/(3n-2)$ and $\omega_o = 0$. Note that as $n \to \infty$, $\omega_\star \to 1/3$. This, together with the fact that experimental studies have used the star topology to test collective estimation in centralized structures [7], motivates our choice of $\omega = 1/3$ in numerical simulations and empirical analysis. In section S4, we show that our results are robust to this choice of $\omega$.

Although our main proposition in section S2.1 gives a lower bound that is valid for any $n$, most of the subsequent analysis concerns the limiting behavior of the collective estimates as $n \to \infty$. Our results extend to networks with a finite (non-increasing) collection of influential agents. To accommodate such cases, one would replace $\mathbf{a}_{1,0}$ in (S.5) with the (possibly weighted) average of the initial estimates of the $k$ influential agents, for some fixed constant $k$. One can similarly consider generalizations where the remaining, non-influential agents have unequal — but all vanishing — weights (going to zero as $n \to \infty$).

## S1.2 $\Omega$: Will the estimation context benefit from centralization?

We consider a case where agents are randomly placed in the social influence network. This is typical of many experimental setups [8, 7, 6]. Let $\mathbb{E}^{\theta}_{\mu,\sigma}$ be the expectation with respect to the random draws of the $n$ i.i.d. initial estimates, $\mathbf{a}_{1,0}, \ldots, \mathbf{a}_{n,0} \sim \mathcal{F}^{\theta}_{\mu,\sigma}$, and let $\mathbb{P}^{\theta}_{\mu,\sigma}$ be the corresponding probability measure. The expected mean square root error, mean absolute error, and mean squared error of the collective estimate are given by:

$$\text{MSRE}_n(\bar{w}, \mathcal{F}^{\theta}_{\mu,\sigma}) := \mathbb{E}^{\theta}_{\mu,\sigma}\left[|\mathbf{a}^n(\bar{w}) - \theta|^{1/2}\right],$$

$$\text{MAE}_n(\bar{w}, \mathcal{F}^{\theta}_{\mu,\sigma}) := \mathbb{E}^{\theta}_{\mu,\sigma}\left[|\mathbf{a}^n(\bar{w}) - \theta|\right],$$

$$\text{MSE}_n(\bar{w}, \mathcal{F}^{\theta}_{\mu,\sigma}) := \mathbb{E}^{\theta}_{\mu,\sigma}\left[(\mathbf{a}^n(\bar{w}) - \theta)^2\right].$$

In order to investigate the interaction between the network structure and the distribution of the initial estimates, (i.e. the *estimation context*: a population of agents performing a particular estimation task), we propose the following measure of how the collective estimate $\mathbf{a}^n(\bar{w}) = \bar{w}^T \bar{\mathbf{a}}_0$ performs against a fully decentralized aggregate $\mathbf{a}^n(0)$:

$$\Omega_n(\bar{w}, \mathcal{F}^{\theta}_{\mu,\sigma}) := \mathbb{P}^{\theta}_{\mu,\sigma}[|\mathbf{a}^n(\bar{w}) - \theta| < |\mathbf{a}^n(0) - \theta|],$$

where $\bar{w}$ is the centrality vector in (S.4). Restricting attention to the class of networks in subsection S1.1, with $\mathbf{a}^n(\omega) = \omega \mathbf{a}_{1,0} + (1-\omega)\frac{1}{n}\sum_{i=1}^{n} \mathbf{a}_{i,0}$, we can write:

$$\Omega_n(\omega, \mathcal{F}^{\theta}_{\mu,\sigma}) = \mathbb{P}^{\theta}_{\mu,\sigma}[|\mathbf{a}^n(\omega) - \theta| < |\mathbf{a}^n(0) - \theta|].$$

This measure, $\Omega_n(\omega, \mathcal{F}^{\theta}_{\mu,\sigma})$, corresponds to the probability that a network with social influence centralization $\omega > 0$ outperforms a decentralized network with $\omega = 0$, in absolute error performance. Similarly, for other performance measures, we can write:

$$\text{MSRE}_n(\omega, \mathcal{F}^{\theta}_{\mu,\sigma}) := \mathbb{E}^{\theta}_{\mu,\sigma}\left[|\mathbf{a}^n(\omega) - \theta|^{1/2}\right], \tag{S.6}$$

$$\text{MAE}_n(\omega, \mathcal{F}^{\theta}_{\mu,\sigma}) := \mathbb{E}^{\theta}_{\mu,\sigma}\left[|\mathbf{a}^n(\omega) - \theta|\right],$$

$$\text{MSE}_n(\omega, \mathcal{F}^{\theta}_{\mu,\sigma}) := \mathbb{E}^{\theta}_{\mu,\sigma}\left[(\mathbf{a}^n(\omega) - \theta)^2\right].$$

Our focus in section S2 is on $\Omega_n$, as the probability of the outcome of interest, to understand if and when the estimation context will benefit from centralization. In section S2, we present a theoretical and numerical analysis of the properties of $\Omega_n$. In particular, we show how the behavior of $\Omega_n$ varies with the estimation context, i.e. the distribution of the initial estimates. In section S2.1, we present a theoretical lower bound on $\Omega_n(\omega, \mathcal{F}_{\mu,\sigma}^\theta)$ and analyze its behavior for various classes of distributions, $\mathcal{F}_{\mu,\sigma}^\theta$. In section S2.2, we supplement these findings by numerical analysis and simulations. We present our numerical results in the main text for $\omega = 1/3$ and $n = 50$. In section S4 we show that our results are robust to our choices of $\omega$ and $n$. Of note, $\Omega_n$ is concerned only with the probability of the following event: the collective estimate generated by the agents interacting in a centralized influence structure will be closer to the truth than the collective estimate generated by the agents in a decentralized structure. This is not the same as comparing the expected loss or error magnitudes. In section S4, Figure S4, we show the results for various loss function choices.

In section S3, we propose a feature of the estimation context that can be empirically measured to predict when social influence improves the collective estimation accuracy in prior empirical studies. The proposed feature, $R$, is based on how heavy-tailed the distribution of the initial estimates is. It is defined as the relative fit (measured in log likelihood) of a log-normal versus a normal distribution to the empirical distribution of the initial estimates. In particular, in experimental conditions with no social interactions, an external observers polls each of the participants for their opinions. Therefore, in the absence of social influence, the aggregate is given by $\mathbf{a}^n(0) = (1/n) \sum_{i=1}^n \mathbf{a}_{i,0}$, which is equivalent to a fully decentralized influence structure. On the other hand, in the presence of social influence, the participants revise their estimates as a result of their social interactions, thus leading to an aggregate that is a weighted average of the initial estimate, $\mathbf{a}^n(\bar{w}) = \sum_{i=1}^n w_i \mathbf{a}_{i,0}$. Hence, social interaction leads to a collective estimate that is less decentralized. In our model, we capture this case by $\mathbf{a}^n(\omega)$ for some $\omega > 0$. Our empirical results show that $R$ can significantly predict when the latter is more accurate than $\mathbf{a}^n(0)$.

# S2 Theoretical analysis of $\Omega$

In this section, we first propose a lower bound on $\Omega_n(\omega, \mathcal{F}_{\mu,\sigma}^\theta)$ in subsection S2.1, followed by analyses of its behavior for different distributions in five sub-subsections: these are the Pareto, S2.1.1, log-Laplace, S2.1.2, log-normal, S2.1.3, other heavy-tailed, S2.1.4, and thin-tailed distributions, S2.1.5. In subsection S2.2, we describe the procedure for direct numerical simulations of $\Omega$ for these various classes and offer additional numerical insights.

## S2.1 The main proposition

Motivated by empirical literature that pose estimation questions to human participants, we focus on $0 < \theta$ and distributions $\mathcal{F}_{\mu,\sigma}^\theta$ with support over positive reals. Fix $0 < \theta$, $0 < \omega < 1$, $\theta/(1-\omega) < \beta$, and consider the event $\mathcal{E}_1 = \{\mathbf{a}_{1,0} < \beta\}$. Note that for many distributions we can make $\mathbb{P}_{\mu,\sigma}^\theta(\mathcal{E}_1)$ arbitrarily close to one by taking $\beta$ large enough. Next consider the event $\mathcal{E}_n = \{\mathbf{a}^n(0) > \beta + \mathbf{a}_{1,0}/n\}$. Note that $\mathcal{E}_n$ implies $\mathbf{a}^n(0) > \beta$; furthermore $\mathcal{E}_1$ and $\mathcal{E}_n$ are independent events. On the other hand, conditioned on the events $\mathcal{E}_1$ and $\mathcal{E}_n$, we have:

$$\begin{aligned}|\mathbf{a}^n(\omega) - \theta| &= |\omega \mathbf{a}_{1,0} + (1-\omega)\mathbf{a}^n(0) - \theta| < \omega \mathbf{a}_{1,0} + |(1-\omega)\mathbf{a}^n(0) - \theta| \\ &< \omega\beta + |(1-\omega)\mathbf{a}^n(0) - \theta| \\ &= \omega\beta + (1-\omega)\mathbf{a}^n(0) - \theta < \omega \mathbf{a}^n(0) + (1-\omega)\mathbf{a}^n(0) - \theta = |\mathbf{a}^n(0) - \theta|,\end{aligned}$$

where in the second line we have used $\beta > \mathbf{a}_{1,0}$, and in the third line we have used $\mathbf{a}^n(0) > \beta$ and $(1-\omega)\mathbf{a}^n(0) > (1-\omega)\beta > \theta$. Hence, conditioned on $\mathcal{E}_1 \cap \mathcal{E}_n$, we have $|\mathbf{a}^n(\omega) - \theta| < |\mathbf{a}^n(0) - \theta|$, i.e., centralized networks outperform decentralized ones. We can thus bound $\Omega_n(\omega, \mathcal{F}_{\mu,\sigma}^\theta)$, the probability that a social influence network with centralization $\omega$ outperforms a decentralized one ($\omega = 0$) in absolute error measure:

$$\Omega_n(\omega, \mathcal{F}_{\mu,\sigma}^\theta) \geq \mathbb{P}_{\mu,\sigma}^\theta[\mathcal{E}_1 \cap \mathcal{E}_n] = \mathbb{P}_{\mu,\sigma}^\theta[\mathcal{E}_1]\mathbb{P}_{\mu,\sigma}^\theta[\mathcal{E}_n]. \quad (S.7)$$

Denoting the cumulative distribution $F^\theta_{\mu,\sigma}(x) := \mathbb{P}^\theta_{\mu,\sigma}[\mathbf{a}_{1,0} \leq x]$ and the tail probability $\bar{F}^\theta_{\mu,\sigma}(x) := \mathbb{P}^\theta_{\mu,\sigma}[\mathbf{a}_{1,0} > x]$, we can write $\mathbb{P}^\theta_{\mu,\sigma}[\mathcal{E}_1] = \mathbb{P}^\theta_{\mu,\sigma}[\mathbf{a}_{1,0} \leq \beta] = F^\theta_{\mu,\sigma}(\beta)$, and

$$\mathbb{P}^\theta_{\mu,\sigma}[\mathcal{E}_n] = \mathbb{P}^\theta_{\mu,\sigma}[\sum_2^n \mathbf{a}_{i,0} > n\beta] \geq \mathbb{P}^\theta_{\mu,\sigma}[\max_{2,\ldots,n} \mathbf{a}_{i,0} > n\beta] = 1 - F^\theta_{\mu,\sigma}(n\beta)^{n-1}. \quad \text{(S.8)}$$

Using $\mathbb{P}^\theta_{\mu,\sigma}[\mathcal{E}_1]\mathbb{P}^\theta_{\mu,\sigma}[\mathcal{E}_n] \geq F^\theta_{\mu,\sigma}(\beta)(1 - F^\theta_{\mu,\sigma}(n\beta)^{n-1})$, together with (S.7), we arrive at our main proposition:

**Proposition.** $\Omega_n(\omega, \mathcal{F}^\theta_{\mu,\sigma}) \geq \sup_{\beta > \theta/(1-\omega)} \left\{ F^\theta_{\mu,\sigma}(\beta)(1 - F^\theta_{\mu,\sigma}(n\beta)^{n-1}) \right\}.$

To proceed, let us denote the lower bound:

$$\underline{\Omega}_n(\omega, \mathcal{F}^\theta_{\mu,\sigma}) := \sup_{\beta > \theta/(1-\omega)} \left\{ F^\theta_{\mu,\sigma}(\beta)(1 - F^\theta_{\mu,\sigma}(n\beta)^{n-1}) \right\}. \quad \text{(S.9)}$$

Some observations are now in order regarding the behavior of the lower bound, $\underline{\Omega}_n(\omega, \mathcal{F}^\theta_{\mu,\sigma})$. First note that $\underline{\Omega}_n(\omega, \mathcal{F}^\theta_{\mu,\sigma})$ is decreasing in $\theta$ and $\omega$. Secondly, the asymptotic behavior of $\underline{\Omega}_n(\omega, \mathcal{F}^\theta_{\mu,\sigma})$, as $n \to \infty$, is determined by $F^\theta_{\mu,\sigma}(n\beta)^{n-1}$. Note that $F^\theta_{\mu,\sigma}(n\beta) \leq 1$ and $F^\theta_{\mu,\sigma}(\beta) \to 1$ as $n \to \infty$ for any $\beta > 0$. Therefore, if $F^\theta_{\mu,\sigma}(n\beta) \to 1$ at a slow enough rate such that $F^\theta_{\mu,\sigma}(n\beta)^{n-1}$ is bounded away from one for all $n$, then $F^\theta_{\mu,\sigma}(\beta)(1 - F^\theta_{\mu,\sigma}(n\beta)^{n-1})$ is bounded away from zero as $n \to \infty$. Subsequently, for various classes of distributions with slow enough tail decay, we can establish nontrivial lower bounds $\underline{\Omega}_n(\omega, \mathcal{F}^\theta_{\mu,\sigma}) > 0$.

In subsections S2.1.1 to S2.1.3, we give conditions on the estimation context, i.e., the distribution of the initial estimates (parameterized by $\mu$, $\sigma$, and $\theta$), of well-known heavy-tailed distributions such that $\underline{\Omega}_n, \Omega_n \to 1$ as $n \to \infty$. In subsection S2.1.4, we discuss the general properties of heavy-tailed distributions that make them relevant to our proposition. Finally, in subsection S2.1.5 we present countervailing arguments for thin-tailed distributions.

### S2.1.1 Pareto (power-law)

Pareto or power-law distributions are archetypal, heavy-tailed distributions characterized by their polynomial tail decay. Consider a Pareto distribution with location parameter $\theta e^\mu$ and

shape parameter $\sigma$, defined as follows:

$$\bar{F}^\theta_{\mu,\sigma}(x) = (\theta e^\mu/x)^{1/\sigma}, \text{ for } x \geq \theta e^\mu. \tag{S.10}$$

For Pareto distributions we have:

$$F^\theta_{\mu,\sigma}(\beta)(1 - F^\theta_{\mu,\sigma}(n\beta)^{n-1}) = \left(1 - \left(\frac{\theta e^\mu}{\beta}\right)^{1/\sigma}\right)\left(1 - \left(1 - \left(\frac{\theta e^\mu}{n\beta}\right)^{1/\sigma}\right)^{n-1}\right)$$

$$\asymp \left(1 - \left(\frac{\theta e^\mu}{\beta}\right)^{1/\sigma}\right)\left(1 - e^{-n(\theta e^\mu/n\beta)^{1/\sigma}}\right), \tag{S.11}$$

where we used the $n \to \infty$ asymptotic equality

$$e^{-n(\theta e^\mu/n\beta)^{1/\sigma}} \asymp \left(1 - \left(\frac{\theta e^\mu}{n\beta}\right)^{1/\sigma}\right)^{n-1}. \tag{S.12}$$

We now consider the three distinct $n \to \infty$ limiting behavior that arise for $\sigma > 1$, $\sigma = 1$, and $\sigma < 1$:

- For $\sigma > 1$, as $n \to \infty$ we get

$$\underline{\Omega}_n(\omega, \mathcal{F}^\theta_{\mu,\sigma}) \geq 1 - (\theta e^\mu/\beta)^{1/\sigma} \text{ for all } \beta > \theta/(1-\omega),$$

and letting $\beta \to \infty$ we conclude that

$$\Omega_n(\omega, \mathcal{F}^\theta_{\mu,\sigma}) \asymp \underline{\Omega}_n(\omega, \mathcal{F}^\theta_{\mu,\sigma}) \asymp 1, \text{ for } \sigma > 1, \text{ and any } 0 < \theta, 0 < \omega < 1, \text{ and real } \mu.$$

- Replacing $\sigma = 1$ in (S.11), the lower bound $\underline{\Omega}_n(\omega, \mathcal{F}^\theta_{\mu,1})$ can be calculated as follow:

$$\sup_{\beta > \theta/(1-\omega)} \left\{1 - \frac{\theta e^\mu}{\beta} - e^{-\theta e^\mu/\beta} + \frac{\theta e^\mu}{\beta} e^{-\theta e^\mu/\beta}\right\}$$

$$= \begin{cases} W_0(e^2) + 1/W_0(e^2) - 2 \approx 0.199, & \text{if } \mu \geq \log\left(\frac{2 - W_0(e^2)}{1 - \omega}\right), \\ 1 - e^\mu(1-\omega) - e^{-e^\mu(1-\omega)} + e^\mu(1-\omega)e^{-e^\mu(1-\omega)}, & \text{if } \mu < \log\left(\frac{2 - W_0(e^2)}{1 - \omega}\right). \end{cases}$$

S9

To see why, denote $x = \theta e^\mu/\beta$. The maximum of $1-x-e^{-x}+xe^{-x}$ occurs at $x^\star$ satisfying $-1 + 2e^{-x^\star} - x^\star e^{-x^\star} = 0$. The latter has a unique solution over the positive reals given by $x^\star = 2 - W_0(e^2) \approx 0.443$, where $W_0$ is the principal branch Lambert $W$ function, uniquely satisfying $W_0(e^2)e^{W_0(e^2)} = e^2$ on positive reals. Replacing $x = 2 - W_0(e^2)$ in $1 - x - e^{-x} + xe^{-x}$ gives the maximum value $W_0(e^2) + 1/W_0(e^2) - 2$ realized at $\beta = \theta e^\mu/(2 - W_0(e^2))$ if $e^\mu \geq (2 - W_0(e^2))/(1 - \omega)$. If $e^\mu < (2 - W_0(e^2))/(1 - \omega)$, then the supremum is achieved with $\beta = \theta/(1 - \omega)$ at a value that is strictly less than $W_0(e^2) + 1/W_0(e^2) - 2 \approx 0.199$.

- For $\sigma < 1$, as $n \to \infty$ we get

$$\left(1 - \left(\frac{\theta e^\mu}{\beta}\right)^{1/\sigma}\right)\left(1 - e^{-n(\theta e^\mu/n\beta)^{1/\sigma}}\right) \to 0.$$

for any $\beta$.

We thus obtain the following asymptotic characterization of the lower bound for Pareto distributions (indicating a phase transition at $\sigma = 1$):

$$\underline{\Omega}_n(\omega, \mathcal{F}^\theta_{\mu,\sigma}) \asymp \begin{cases} 0, & \text{if } \sigma < 1, \\ (1 - e^\mu(1 - \omega))(1 - e^{-e^\mu(1-\omega)}), & \text{if } \sigma = 1 \text{ and } \mu < \log\left(\frac{2-W_0(e^2)}{1-\omega}\right), \\ W_0(e^2) + 1/W_0(e^2) - 2 \approx 0.199, & \text{if } \sigma = 1 \text{ and } \mu \geq \log\left(\frac{2-W_0(e^2)}{1-\omega}\right), \\ 1, & \text{if } \sigma > 1. \end{cases}$$
(S.13)

In Figure S1.A, top, we have plotted (S.13) for $\omega = 1/3$, $\theta = 2$, and $n = 50$. Comparing with the direct numerical simulation in Figure S1.A, bottom, shows how the bound gets tighter for large $\sigma$.

### S2.1.2 Log-Laplace

Jayles et al. [10] point out that log-Laplace provides a better fit to the empirically measured distribution of the initial estimates, compared to log-Cauchy [11], or log-normal. Here, we analyze the asymptotic behavior of the proposed lower bound as $n \to \infty$, when the initial estimates are distributed according to a log-Laplace distribution with parameters $\log \theta + \mu$ and

$\sigma$:

$$F_{\mu,\sigma}^{\theta}(x) = \begin{cases} \frac{1}{2}\exp\left(\frac{\log(x/\theta)-\mu}{\sigma}\right), & \text{if } \log(x/\theta) < \mu, \\ 1 - \frac{1}{2}\exp\left(\frac{\mu-\log(x/\theta)}{\sigma}\right), & \text{if } \log(x/\theta) \geq \mu. \end{cases} \quad (S.14)$$

For $n$ large enough, we have $\log(n\beta/\theta) \geq \mu$, and using the same asymptotics as in (S.12) we get:

$$\begin{aligned} F_{\mu,\sigma}^{\theta}(n\beta)^{n-1} &= \left(1 - \frac{1}{2}\exp\left(\frac{\mu - \log(n\beta/\theta)}{\sigma}\right)\right)^{n-1} \\ &= \left(1 - \left(\frac{1}{n}\right)^{1/\sigma}\exp\left(\frac{\mu + \log\theta - \log\beta}{\sigma} - \log(2)\right)\right)^{n-1} \\ &\asymp \exp\left(-n\left(\frac{1}{n}\right)^{1/\sigma}\exp\left(\frac{\mu + \log\theta - \log\beta}{\sigma} - \log(2)\right)\right) \\ &\asymp \begin{cases} 1, & \text{if } \sigma < 1, \\ \exp\left(\frac{-e^{\mu}\theta}{2\beta}\right), & \text{if } \sigma = 1, \\ 0, & \text{if } \sigma > 1. \end{cases} \end{aligned}$$

Subsequently, for $\sigma < 1$ we have $F_{\mu,\sigma}^{\theta}(\beta)(1 - F_{\mu,\sigma}^{\theta}(n\beta)^{n-1}) \asymp 0$, for any $\beta$. On the other hand, for $\sigma > 1$ we have $F_{\mu,\sigma}^{\theta}(\beta)(1 - F_{\mu,\sigma}^{\theta}(n\beta)^{n-1}) \asymp F_{\mu,\sigma}^{\theta}(\beta)$, which is increasing in $\beta$ and goes to one as $\beta$ increases to $\infty$. Hence,

$$\underline{\Omega}_n(\omega, \mathcal{F}_{\mu,\sigma}^{\theta}) = \sup_{\beta > \theta/(1-\omega)} \left\{ F_{\mu,\sigma}^{\theta}(\beta)(1 - F_{\mu,\sigma}^{\theta}(n\beta)^{n-1}) \right\} \asymp 1, \text{ for } \sigma > 1.$$

Finally, for $\sigma = 1$, we have:

$$F_{\mu,1}^{\theta}(\beta)(1 - F_{\mu,1}^{\theta}(n\beta)^{n-1}) \asymp \begin{cases} \left(1 - \exp\left(\frac{-e^{\mu}\theta}{2\beta}\right)\right) \frac{\beta}{2\theta e^{\mu}}, & \text{if } \frac{\beta}{e^{\mu}\theta} < 1, \\ \left(1 - \exp\left(\frac{-e^{\mu}\theta}{2\beta}\right)\right)\left(1 - \frac{e^{\mu}\theta}{2\beta}\right), & \text{if } \frac{\beta}{e^{\mu}\theta} \geq 1. \end{cases}$$

Optimizing $\beta$ gives:

$$\underline{\Omega}_n(\omega, \mathcal{F}_{\mu,1}^\theta) \asymp \begin{cases} \frac{1-1/\sqrt{e}}{2} \approx 0.1967, & \text{if } \mu \geq -\log(1-\omega), \\ \left(1 - e^{-e^\mu(1-\omega)/2}\right)(1 - e^\mu(1-\omega)/2) < \frac{1-1/\sqrt{e}}{2}, & \text{if } \mu < -\log(1-\omega). \end{cases}$$

We summarize the above results in the following asymptotic characterization of the lower bound for log-Laplace distributions, with a phase transition at $\sigma = 1$:

$$\underline{\Omega}_n(\omega, \mathcal{F}_{\mu,\sigma}^\theta) \asymp \begin{cases} 0, & \text{if } \sigma < 1, \\ \left(1 - e^{-e^\mu(1-\omega)/2}\right)(1 - e^\mu(1-\omega)/2), & \text{if } \sigma = 1 \text{ and } \mu < -\log(1-\omega), \\ \frac{1-1/\sqrt{e}}{2} \approx 0.1967, & \text{if } \sigma = 1 \text{ and } \mu \geq -\log(1-\omega), \\ 1, & \text{if } \sigma > 1. \end{cases} \quad \text{(S.15)}$$

In Figure S1.B, top, we have plotted (S.15) for $\omega = 1/3$, $\theta = 2$, and $n = 50$. Comparing with the direct numerical simulation in Figure S1.B, bottom, shows how the bound gets tighter for large $\sigma$.

### S2.1.3 Log-normal

Several empirical studies report a log-normal distribution for the initial estimates [7, 8, 12]. Here, we analyze the case where the initial estimates are distributed according to a log-normal distribution with parameters $\log \theta + \mu$ and $\sigma$:

$$F_{\mu,\sigma}^\theta(x) = \Phi\left(\frac{\log(x/\theta) - \mu}{\sigma}\right), \, x > 0, \quad \text{(S.16)}$$

where $\Phi$ is the standard normal distribution. We next apply the following control over the Gaussian tail:

$$1 - \frac{1}{\sqrt{2\pi}t}e^{-t^2/2} \leq \Phi(t) \leq 1 - \frac{t}{\sqrt{2\pi}(t^2+1)}e^{-t^2/2}$$

S12

to obtain:

$$\left(1 - \frac{\sigma}{\sqrt{2\pi}(\log(n\beta/\theta) - \mu)} \exp\left(-\frac{(\log(n\beta/\theta) - \mu)^2}{2\sigma^2}\right)\right)^{n-1} \leq F_{\mu,\sigma}^{\theta}(n\beta)^{n-1}$$
$$\leq \left(1 - \frac{(\log(n\beta/\theta) - \mu)\sigma}{\sqrt{2\pi}((\log(n\beta/\theta) - \mu)^2 + \sigma^2)} \exp\left(-\frac{(\log(n\beta/\theta) - \mu)^2}{2\sigma^2}\right)\right)^{n-1}. \quad \text{(S.17)}$$

We next choose $\sigma = \sigma_n$, with $\beta$ and $\theta$, fixed such that $(\log(n\beta/\theta) - \mu)/\sigma_n \to \infty$ as $n \to \infty$. Taking the limit $n \to \infty$, with $\sigma = \sigma_n$ in (S.17), we get:

$$F_{\mu,\sigma_n}^{\theta}(n\beta)^{n-1} \quad \text{(S.18)}$$
$$\asymp \exp\left(-\frac{n\sigma_n}{\sqrt{2\pi}\log n} \exp\left(-\frac{(\log(n\beta/\theta) - \mu)^2}{2\sigma_n^2}\right)\right)$$
$$= \exp\left(-\frac{1}{\sqrt{2\pi}\log n} \exp\left(\log n + \log \sigma_n - \frac{(\log(n\beta/\theta) - \mu)^2}{2\sigma_n^2}\right)\right)$$

Focusing on the second exponent, we obtain :

$$f_n := \log n + \log \sigma_n - \frac{(\log(n\beta/\theta) - \mu)^2}{2\sigma_n^2} \quad \text{(S.19)}$$
$$= \log n + \log \sigma_n - \frac{\log(n\beta/\theta)^2}{2\sigma_n^2} - \frac{\mu^2}{2\sigma_n^2} + \frac{\mu \log(n\beta/\theta)}{\sigma_n^2}$$
$$= \log n + \log \sigma_n - \frac{(\log n)^2}{2\sigma_n^2} - \frac{\log(\beta/\theta)\log n}{\sigma_n^2} - \frac{\log(\beta/\theta)^2}{2\sigma_n^2} - \frac{\mu^2}{2\sigma_n^2} + \frac{\mu \log(n\beta/\theta)}{\sigma_n^2}.$$

Setting $\sigma_n = \sqrt{\log n / k_n}$ in (S.19) yields:

$$f_n = (1 - k_n/2)\log n + (1/2)\log\log n + (\mu - \log(\beta/\theta))k_n - (1/2)\log k_n$$
$$+ (-(1/2)\log(\beta/\theta)^2 - \mu^2/2 + \mu\log(\beta/\theta))\frac{k_n}{\log n}$$
$$\asymp \begin{cases} \frac{(2-k_n)}{2}\log n, & \text{if } k_n < 2, \\ (1/2)\log\log n, & \text{if } k_n = 2, \\ \frac{(2-k_n)}{2}\log n, & \text{if } k_n > 2. \end{cases}$$

S13

Finally, substituting in (S.18), we get:

$$F^\theta_{\mu,\sigma_n}(n\beta)^{n-1} \asymp \exp\left(-e^{f_n}/(\sqrt{2\pi}\log n)\right) \asymp \begin{cases} 0, & \text{if } \sigma_n > \sqrt{\log n/2}, \\ 1, & \text{if } \sigma_n \leq \sqrt{\log n/2}. \end{cases}$$

The latter together with the fact that $F^\theta_{\mu,\sigma_n}(\beta) = \Phi\left((\log(\beta/\theta) - \mu)/\sigma_n\right) \asymp \Phi(0) = 1/2$, for $\sigma_n > \sqrt{\log n/2}$, leads to the following asymptotic characterization for log-normal distributions (indicating a phase transition at $\sigma_n = \sqrt{\log n/2}$):

$$F^\theta_{\mu,\sigma_n}(\beta)(1 - F^\theta_{\mu,\sigma_n}(n\beta)^{n-1}) \asymp \begin{cases} 0, & \text{if } \sigma_n \leq \sqrt{\log n/2}, \\ 1/2, & \text{if } \sigma_n > \sqrt{\log n/2}, \end{cases}$$

which is true for all $\beta$, and in particular yields:

$$\underline{\Omega}_n(\omega, \mathcal{F}^\theta_{\mu,\sigma}) = \sup_{\beta > \theta/(1-\omega)} \left\{ F^\theta_{\mu,\sigma}(\beta)(1 - F^\theta_{\mu,\sigma}(n\beta)^{n-1}) \right\}$$
$$\asymp \begin{cases} 0, & \text{if } \sigma_n \leq \sqrt{\log n/2}, \\ 1/2, & \text{if } \sigma_n > \sqrt{\log n/2}. \end{cases} \quad (S.20)$$

In Figure S1.C, top, we have plotted (S.20) for $\omega = 1/3$, $\theta = 2$, and $n = 50$. Comparing with the result of the direct numerical simulation, Figure S1.C, bottom, shows how the bound gets tighter for larger $\sigma$.

### S2.1.4 Other heavy-tailed distributions

Many empirical studies [8, 11, 12, 13, 14] point out a heavy-tailed distribution for the numerical estimates (with a few estimates that fall on a fat right tail). Following the proof of the main proposition, we pointed out that for heavy-tailed distributions where $\bar{F}^\theta_{\mu,\sigma}(n\beta)$ decreases slowly, we can provide non-trivial lower bounds on $\Omega_n$ that remain bounded away from zero, $\underline{\Omega}_n(\omega, \mathcal{F}^\theta_{\mu,\sigma}) > 0$, even as $n \to \infty$. In fact, if $\bar{F}^\theta_{\mu,\sigma}(n\beta)$ decreases at a rate that is slower than $1/n$, i.e., $n\bar{F}^\theta_{\mu,\sigma}(n\beta) \to \infty$, then $F^\theta_{\mu,\sigma}(n\beta)^{n-1} \to 0$ as $n \to \infty$. For such slowly decaying tails, the supremum in (S.9) is achieved as $\beta \to \infty$, and we can guarantee that $\underline{\Omega}_n \asymp \Omega_n \asymp 1$; hence, the proposed lower bound is asymptotically tight.

Here, we identify a second way, in which, our proposed lower bound is tighter for heavy-tailed distributions. To this end, let us revisit (S.8) — a critical step in deriving the proposed lower bound:

$$\mathbb{P}^{\theta}_{\mu,\sigma}[\mathcal{E}_n] = \mathbb{P}^{\theta}_{\mu,\sigma}[\sum_{i=2}^{n} \mathbf{a}_{i,0} > n\beta] > \mathbb{P}^{\theta}_{\mu,\sigma}[\max_{2,\ldots,n} \mathbf{a}_{i,0} > n\beta].$$

This inequality is at the heart of the so-called "catastrophe principle" [15, Chapter 3] that applies to many heavy-tailed distributions. Intuitively, this principle entails that when one observes a larger than expected average value for a collection of heavy-tailed random variables, then this observation is most likely explained by the existence of a very large sample in the collection, i.e. a "catastrophe". On the other hand, the countervailing explanation in the case of light-tailed random variables is that "most" of the samples in the collection happen to be larger than expected. Formally, the distribution $\mathcal{F}^{\theta}_{\mu,\sigma}$ of the initial estimates is said to satisfy the catastrophe principle [15, Definition 3.1], if for any $n$:

$$\lim_{t \to \infty} \frac{\mathbb{P}^{\theta}_{\mu,\sigma}[\max_{1,\ldots,n} \mathbf{a}_{i,0} > t]}{\mathbb{P}^{\theta}_{\mu,\sigma}[\sum_{i=1}^{n} \mathbf{a}_{i,0} > t]} = 1.$$

The preceding condition is equivalent to having:

$$\lim_{t \to \infty} \frac{n\mathbb{P}^{\theta}_{\mu,\sigma}[\mathbf{a}_{1,0} > t]}{\mathbb{P}^{\theta}_{\mu,\sigma}[\sum_{i=1}^{n} \mathbf{a}_{i,0} > t]} = 1, \text{ for all } n \geq 2.$$

The latter is the defining property for the subexponential family of distributions, which include many common classes of heavy-tailed distributions such as those considered in subsections S2.1.1 to S2.1.3. Setting $t = n\beta$ and letting $n \to \infty$, we obtain that if $\mathcal{F}^{\theta}_{\mu,\sigma}$ is a member of the subexponential family, then

$$\mathbb{P}^{\theta}_{\mu,\sigma}[\mathcal{E}_n] = \mathbb{P}^{\theta}_{\mu,\sigma}[\sum_{i=2}^{n} \mathbf{a}_{i,0} > n\beta] \asymp \mathbb{P}^{\theta}_{\mu,\sigma}[\max_{2,\ldots,n} \mathbf{a}_{i,0} > n\beta].$$

Hence, for such distributions belonging to the subexponential family our proposed lower bound is asymptotically tight in as much as $\mathbb{P}^{\theta}_{\mu,\sigma}[\mathcal{E}_1 \cap \mathcal{E}_n] \asymp F^{\theta}_{\mu,\sigma}(\beta)(1 - F^{\theta}_{\mu,\sigma}(n\beta)^{n-1})$, and the only way in which our lower bound may be loose is through (S.7), i.e. if $\Omega_n(\omega, \mathcal{F}^{\theta}_{\mu,\sigma}) > \mathbb{P}^{\theta}_{\mu,\sigma}[\mathcal{E}_1 \cap \mathcal{E}_n]$ for all $\beta > \theta/(1-\omega)$ as $n \to \infty$.

It is worth noting that many light-tailed distributions portray an opposite picture, referred to as "conspiracy principle" in [15, Definition 3.2]; formally defined as follows:

$$\lim_{t \to \infty} \frac{\mathbb{P}_{\mu,\sigma}^{\theta}[\max_{1,\ldots,n} \mathbf{a}_{i,0} > t]}{\mathbb{P}_{\mu,\sigma}^{\theta}[\sum_{i=1}^{n} \mathbf{a}_{i,0} > t]} = 0, \text{ for all } n \geq 2.$$

As an example, suppose that the initial estimates are exponentially distributed with mean $\theta e^{\mu}$ and the following tail probability:

$$\bar{F}_{\mu,\sigma}^{\theta}(x) = e^{-x/\theta e^{\mu}}, x > 0.$$

Then their sum follows an Erlang distribution, satisfying:

$$\mathbb{P}_{\mu,\sigma}^{\theta}[\sum_{2}^{n} \mathbf{a}_{i,0} > n\beta] = \sum_{k=0}^{n-2} \frac{e^{-n\beta/(\theta e^{\mu})}}{k!} \left(\frac{n\beta}{\theta e^{\mu}}\right)^{k},$$

such that

$$\mathbb{P}_{\mu,\sigma}^{\theta}[\max_{2,\ldots,n} \mathbf{a}_{i,0} > n\beta] \asymp (n-1)\mathbb{P}_{\mu,\sigma}^{\theta}[\mathbf{a}_{1,0} > n\beta] = (n-1)e^{-n\beta/(\theta e^{\mu})}$$

$$\ll \sum_{k=0}^{n-2} \frac{e^{-n\beta/(\theta e^{\mu})}}{k!} \left(\frac{n\beta}{\theta e^{\mu}}\right)^{k} = \mathbb{P}_{\mu,\sigma}^{\theta}[\sum_{i=2}^{n} \mathbf{a}_{i,0} > n\beta].$$

and

$$\frac{\mathbb{P}_{\mu,\sigma}^{\theta}[\max_{2,\ldots,n} \mathbf{a}_{i,0} > n\beta]}{\mathbb{P}_{\mu,\sigma}^{\theta}[\sum_{i=2}^{n} \mathbf{a}_{i,0} > n\beta]} \asymp 0.$$

**S2.1.5 Distributions with strong tail decay and classic accounts of the wisdom of crowds**

It is instructive to investigate the behavior of the lower bound for light-tailed distributions as well. Sub-Gaussian distributions are a class of probability distribution with strong tail decay (at least as fast as a Gaussian). Suppose $\mathbf{x}$ is a random variable with mean $\mu + \theta$ and cumulative distribution $F_{\mu,\sigma}^{\theta}$. Furthermore, suppose that $\mathbf{x} - \mu - \theta$ is sub-Gaussain with variance-proxy

parameter $\sigma$, thereby, satisfying:

$$\bar{F}^{\theta}_{\mu,\sigma}(n\beta) = \mathbb{P}^{\theta}_{\mu,\sigma}[\mathbf{x} > n\beta] \leq e^{-(n\beta-\mu-\theta)^2/2\sigma^2}$$

On the other hand, we have $(1 + F^{\theta}_{\mu,\sigma}(n\beta) + F^{\theta}_{\mu,\sigma}(n\beta)^2 + \ldots + F^{\theta}_{\mu,\sigma}(n\beta)^{n-2}) \leq n - 1$, which we can combine with the above to get that

$$F^{\theta}_{\mu,\sigma}(\beta)(1 - F^{\theta}_{\mu,\sigma}(n\beta)^{n-1})$$
$$= F^{\theta}_{\mu,\sigma}(\beta)(1 - F^{\theta}_{\mu,\sigma}(n\beta))(1 + F^{\theta}_{\mu,\sigma}(n\beta) + F^{\theta}_{\mu,\sigma}(n\beta)^2 + \ldots + F^{\theta}_{\mu,\sigma}(n\beta)^{n-2})$$
$$\leq nF^{\theta}_{\mu,\sigma}(\beta) \exp\left(-\frac{(n\beta-\mu-\theta)^2}{2\sigma^2}\right) \to 0, \text{ as } n \to \infty, \text{ for any } \beta > 0.$$

Therefore, there are no set of parameters $\mu$ and $\sigma$ that lead to a non-trivial, asymptotic lower bound on $\Omega_n$ for random variables with sub-Gaussian tails: $\underline{\Omega}_n(\omega, \mathcal{F}^{\theta}_{\mu,\sigma}) \asymp 0$, for all $\theta$, $\mu$, $\sigma$. As an example, consider the folded Gaussian distribution, which is defined as the absolute value of a normally distributed random variable with mean $\theta e^{\mu}$ and variance $\sigma$:

$$F^{\theta}_{\mu,\sigma}(x) = \Phi\left(\frac{x - \theta e^{\mu}}{\sigma}\right) + \Phi\left(\frac{x + \theta e^{\mu}}{\sigma}\right) - 1, x > 0, \tag{S.21}$$

where $\Phi$ is the standard normal distribution. In Figure S1.D, bottom, we have plotted $\Omega_n(\omega, \mathcal{F}^{\theta}_{\mu,\sigma})$ with $\omega = 1/3$, $\theta = 2$, $n = 50$, and initial estimates following a folded-Gaussian distribution. There are no range of distribution parameters, $\mu$ and $\sigma$, for which $\Omega_n$ increases above 0.6. Indeed, for such light-tailed distributions, admitting finite first and second moments, we can show that the limiting expected absolute error of the collective estimate with centralization $\omega$, $\mathbf{a}^n(\omega)$, is higher than the decentralized baseline, $\mathbf{a}^n(0)$.

Consider the case where $\mathcal{F}^{\theta}_{\mu,\sigma}$ admits the following finite first and second moments: $\mathbb{E}^{\theta}_{\mu,\sigma}[\mathbf{a}_{1,0}] = \theta e^{\mu}$, and $\mathbb{E}^{\theta}_{\mu,\sigma}[\mathbf{a}^2_{1,0}] = \theta^2 e^{2\mu} + \sigma^2$. Then $\mathbf{a}^n(0) \to \theta e^{\mu}$, and $\mathbf{a}^n(\omega) \to \omega \mathbf{a}_{1,0} + (1-\omega)\theta e^{\mu}$, both almost surely, as $n \to \infty$. Hence,

$$\mathbb{E}^{\theta}_{\mu,\sigma}[|\mathbf{a}^n(\omega) - \theta|] > |\mathbb{E}^{\theta}_{\mu,\sigma}[\mathbf{a}^n(\omega) - \theta]| = |\theta(1 - e^{\mu})| \asymp \mathbb{E}^{\theta}_{\mu,\sigma}[|\mathbf{a}^n(0) - \theta|].$$

We can repeat the same calculations for the expected mean squared errors as well:

$$\mathbb{E}^{\theta}_{\mu,\sigma}[(\mathbf{a}^n(0) - \theta)^2] \to \theta^2(1 - e^\mu)^2, \text{ and}$$

$$\mathbb{E}^{\theta}_{\mu,\sigma}[(\mathbf{a}^n(\omega) - \theta)^2] \to \theta^2(1 - e^\mu)^2 + \omega^2\sigma^2 > \theta^2(1 - e^\mu)^2 \asymp \mathbb{E}^{\theta}_{\mu,\sigma}[(\mathbf{a}^n(0) - \theta)^2].$$

For distributions with light tails decentralized networks outperform centralized ones, in expectation for absolute and squared errors, for any choice of parameters $\mu$ and $\sigma$. This verifies the classical accounts of the wisdom of crowds, whereby the law of large numbers guarantees almost sure convergence of the collective estimate for structures with vanishing individual influences [16, Proposition 2].

Indeed, among all convex combinations the simple average, $\mathbf{a}^n(0)$, has the minimum variance. Since all estimators in this class, $\mathbf{a}^n(\omega), \omega \geq 0$, have the same expected value, the simple average, $\mathbf{a}^n(0)$, is, in fact, the mean squared error (MSE) minimizer in this class. In this subsection, we point out that even though the variance/MSE of $\mathbf{a}^n(\omega)$ is minimum when $\omega = 0$, $\mathbf{a}^n(\omega)$, with $\omega > 0$ fixed, can "often" outperform $\mathbf{a}^n(0)$, i.e., fall closer to the truth, $\theta$. We show this by lower bounding the probability, $\Omega_n$, that $|\mathbf{a}^n(\omega) - \theta| < |\mathbf{a}^n(0) - \theta|$. Subsequently, we identify heavy-tailedness conditions that make this event likely and $\Omega_n$ large.

It is worth highlighting that although for $\mu = 0$, the sample mean (simple average) is a minimum-variance, unbiased estimator (MVUE); in statistics, it is well-known that simple average is not "optimal" for closeness in many situations. For instance, in the presence of outliers or extreme values the sample median is preferred to sample mean due to its robustness properties [17]. Our calculations in Subsections S2.1.1-S2.1.4 are of a similar flavor, pointing out the superiority of a weighted average when the underlying distributions are heavy-tailed.

In subsection S2.2, we discuss direct numerical simulation of the value of $\Omega_n$ for various distribution classes. In sub-subsection S2.2.1, we identify other comparable right-skewness conditions for making $\Omega_n$ large, e.g., greater than $1/2$, by analyzing the location of the medians of the two estimators, $\mathbf{a}^n(\omega), \omega > 0$ and $\mathbf{a}^n(0)$. In section S3, we show that the feature, $\Omega$, that we identify from this theory has significant explanatory power for determining whether experimentally measured collective estimation outcomes improve, after group members interact with each other.

## S2.2 Numerical simulations

For numerical simulations, we have fixed $\theta = 2$, $\omega = 1/3$, and $n = 50$. The choice of $\omega = 1/3$ is arbitrary and our conclusions remain valid for $\omega > 0$, as verified by the robustness checks in section S4. This choice is motivated by our observation in subsection S1.1 that $\omega$ for a star network converges to $1/3$ as $n \to \infty$. This also allows us to juxtapose our simulations with common experimental setups that use the star topology as archetypal of centralized structures [7].

Note that with $\omega = 1/3$ fixed, the dependence of $\Omega_n$ on the network structure is removed. Therefore, $\Omega_n(\omega, \mathcal{F}^\theta_{\mu,\sigma})$ is entirely determined by the distribution of the initial estimates, $\mathcal{F}^\theta_{\mu,\sigma}$, i.e. the estimation context. Here, we study our proposed task feature, $\Omega_n$, numerically for a palette of empirically relevant distributions.

For any distribution of the initial estimates, $\mathcal{F}^\theta_{\mu,\sigma}$, and number of agents, $n$, we calculate $\Omega_n$ using a Monte Carlo method. We sample $n$ initial estimates and calculate the collective estimates, $\mathbf{a}^n(1/3)$ and $\mathbf{a}^n(0)$, using equation (S.5). If $\mathbf{a}^n(1/3)$ is closer to the truth, $\theta$, than $\mathbf{a}^n(0)$, implying that a centralized network performed better than a decentralized network, then we add to our tally of $\Omega_n$. We repeat this procedure $N$ times, where $N$ is large enough to allow for the value of $\Omega_n$ to converge (see the simulation procedure 1). The results in Figure S1 are obtained in this manner with $\theta = 2$, $n = 50$, and $N = 10,000$ for four different distributions: Pareto (S.10), log-Laplace (S.14), log-normal (S.16), and folded-Gaussian (S.21).

---

**Procedure 1:** $\Omega_n(\omega, \mathcal{F}^\theta_{\mu,\sigma})$ computation

---

**Input:** $\mathcal{F}_\theta(\mu, \sigma)$, $n$, $\omega$, $N$
**Output:** $\Omega_n(\omega, \mathcal{F}^\theta_{\mu,\sigma})$
Initialize $\Omega_n(\omega, \mathcal{F}^\theta_{\mu,\sigma}) = 0$.
**for** $j = 1 : N$ **do**
    Sample $\mathbf{a}_{1,0}, \ldots, \mathbf{a}_{n,0} \sim \mathcal{F}_\theta(\mu, \sigma)$.
    Compute $\mathbf{a}^n(0)$ and $\mathbf{a}^n(\omega)$ according to (S.5).
    **if** $|\mathbf{a}^n(\omega) - \theta| < |\mathbf{a}^n(0) - \theta|$ **then**
        Update $\Omega_n(\omega, \mathcal{F}^\theta_{\mu,\sigma}) = \Omega_n(\omega, \mathcal{F}^\theta_{\mu,\sigma}) + 1/N$.
    **end**
**end**

---

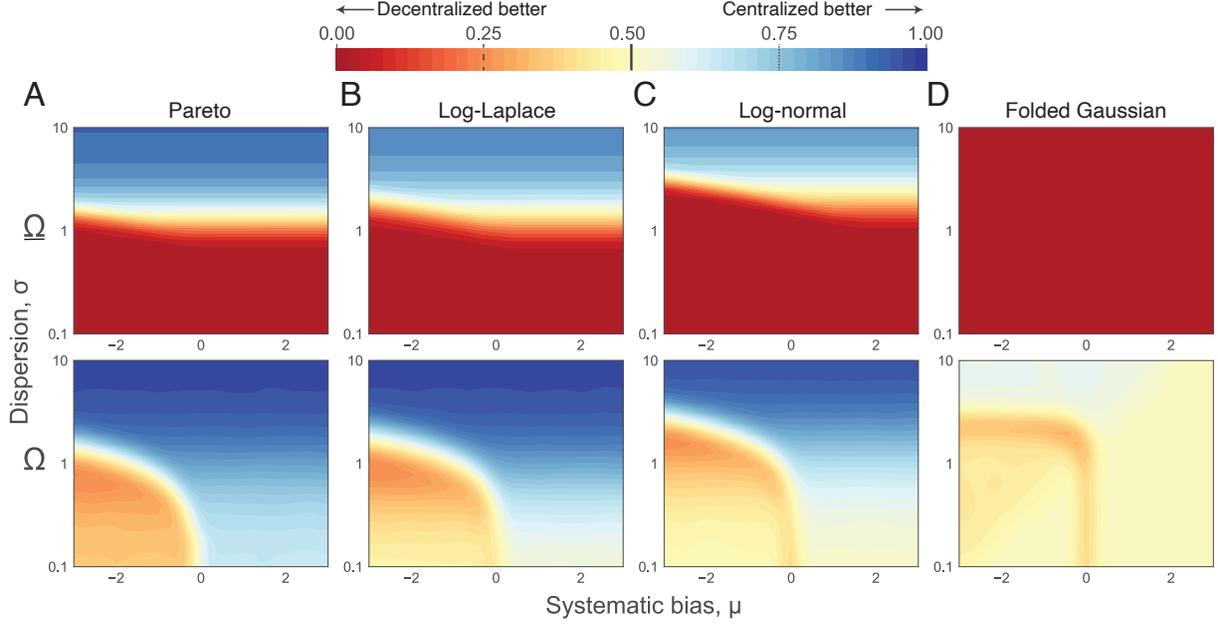

Figure S1: Simulating the lower bound $\underline{\Omega}_n(\omega, \mathcal{F}^\theta_{\mu,\sigma})$ and the actual value $\Omega_n(\omega, \mathcal{F}^\theta_{\mu,\sigma})$, under different distributions for the initial estimates: Pareto (S.10), log-Laplace (S.14), log-normal (S.16), and folded Gaussian (S.21). In all the plots, we have fixed $\omega = 1/3$, $\theta = 2$, and $n = 50$.

### S2.2.1 The effect of the systematic bias, $\mu$

To see the effect of the log-normal distribution parameters, $\mu$ and $\sigma$, in a different light, it is instructive to study the behavior of the median of the random variable $\mathbf{a}^n(\omega)$. In particular, we are interested in the location of Median$[\mathbf{a}^n(\omega)]$ with respect to the truth $\theta$, as the distribution parameter $\mu$ is varied. We do so in the limit of large group sizes, $n \to \infty$. Note that since log-normal distributions have finite moments, the strong law of large numbers applies. Hence, as $n \to \infty$, $\mathbf{a}^n(0)$ converges almost surely to

$$\mathbb{E}[\mathbf{a}_{i,0}] = \exp(\log \theta + \mu + \sigma^2/2)$$

In particular we also have that

$$\lim_{n \to \infty} \text{Median}[\mathbf{a}^n(0)] = \exp(\log \theta + \mu + \sigma^2/2).$$

On the other hand, note that for $\omega > 0$, $\mathbf{a}^n(\omega) = \omega \mathbf{a}_{1,0} + (1-\omega)\mathbf{a}^n(0)$. Hence, as $n \to \infty$, $\mathbf{a}^n(\omega) \to \omega \mathbf{a}_{1,0} + \exp(\log\theta + \mu + \sigma^2/2)$, almost surely. Therefore,

$$\lim_{n\to\infty} \text{Median}[\mathbf{a}^n(\omega)] = \omega \, \text{Median}[\mathbf{a}_{i,0}] + (1-\omega)\exp(\log\theta + \mu + \sigma^2/2)$$
$$= \omega \exp(\log\theta + \mu) + (1-\omega)\exp(\log\theta + \mu + \sigma^2/2),$$

where we have used the fact that $\text{Median}[\mathbf{a}_{i,0}] = \exp(\log\theta + \mu)$. Next, we note that depending on where the distributional parameters, $\theta$, $\sigma$, and $\mu$, are located, three cases may arise:

- If $\exp(\log\theta + \mu + \sigma^2/2) < \theta$, or equivalently, $\mu < -\sigma^2/2$, then $\lim_{n\to\infty} \text{Median}[\mathbf{a}^n(\omega)] < \lim_{n\to\infty} \text{Median}[\mathbf{a}^n(0)] < \theta$. In this case, as $n \to \infty$, at least half of the time, $\mathbf{a}^n(0)$ falls closer to $\theta$ than $\mathbf{a}^n(\omega)$, hence, $\lim_{n\to\infty} \Omega_n < 1/2$; see Figure S2.A.

- If $\exp(\log\theta + \mu) < \theta < \exp(\log\theta + \mu + \sigma^2/2)$, or equivalently, $-\sigma^2/2 < \mu < 0$, then $\lim_{n\to\infty} \text{Median}[\mathbf{a}^n(\omega)] < \theta < \lim_{n\to\infty} \text{Median}[\mathbf{a}^n(0)]$. In this case, the limiting value of $\Omega_n$ may be less then, or greater than, but is close to $1/2$; see Figure S2.B.

- If $\theta < \exp(\log\theta + \mu + \sigma^2/2)$, or equivalently, $0 < \mu$, then $\theta < \lim_{n\to\infty} \text{Median}[\mathbf{a}^n(\omega)] < \lim_{n\to\infty} \text{Median}[\mathbf{a}^n(0)]$. In this case, as $n \to \infty$, at least half of the time, $\mathbf{a}^n(\omega)$ falls closer to $\theta$ than $\mathbf{a}^n(0)$, hence, $\lim_{n\to\infty} \Omega_n > 1/2$; see Figure S2.C.

Finally, it is worth noting that a similar argument applies to any right-skewed and heavy-tailed distribution, for which the population mean exists and is greater than the population median.

## S3 Empirical analysis of estimation contexts in prior work

To empirically illustrate the explanatory power of this theory, we use data from four published experiments [8, 18, 7, 6], in which $2,885$ participants organized into $99$ independent groups completed a total of $54$ estimation tasks generating $15,562$ individual estimations and $687$ collective estimations. Each task induces a different distribution on the initial estimates that are measured empirically. Therefore, each task constitutes an estimation context in our framework and we have a total of $54$ estimation contexts.

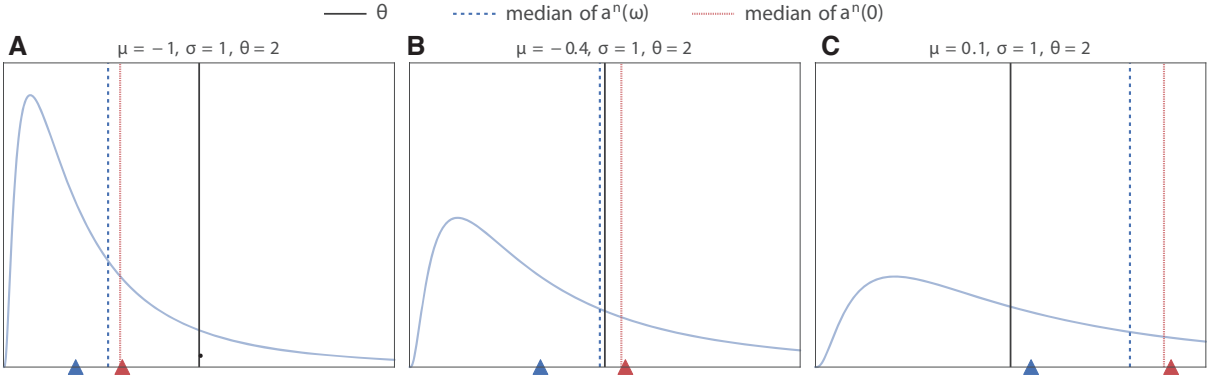

Figure S2: Simulating the medians of $\mathbf{a}^n(\omega)$ and $\mathbf{a}^n(0)$ for three different values of the systematic bias, $\mu$. The distribution median, marked in blue on the x-axis, lags the distribution mean, marked in red. Subsequently, the median of $\mathbf{a}^n(\omega)$ is always less than the median of $\mathbf{a}^n(0)$ for the distributions studied here. In this framing, there are three different levels of bias: panel **A**, when the distribution of initial estimates significantly under-estimates the truth, $\mu < -\sigma^2/2$, then the median of $\mathbf{a}^n(0)$ is closer to the truth, $\theta$, than the median of $\mathbf{a}^n(\omega)$, in this case, $\Omega_n < 1/2$; panel **B**, when the distribution of initial estimates slightly under-estimates the truth, $-\sigma^2/2 < \mu < 0$, then the truth lies between the medians of $\mathbf{a}^n(\omega)$ and $\mathbf{a}^n(0)$, and $\Omega \approx 0.5$; panel **C**, when the distribution of initial estimates over-estimates the truth, $\mu > 0$, then the median $\mathbf{a}^n(\omega)$ is closer to the truth, leading to $\Omega_n > 1/2$. In these simulations, we have fixed $\omega = 1/3$, $\theta = 2$, $n = 50$, and $N = 10,000$, where $N$ is the number of samples used to simulate the median values numerically.

## S3.1 Regression Analysis

Our main empirical analyses, shown in Figure 3 and Table S1, are based on a mixed effect model with a random effect to account for the nested structure of the data.

In particular, the logistic regression equation for Figure 3.C (Table S1, Model 1) is:

$$y_{ij} = \frac{1}{1 + \exp(\beta_0 + \beta_1 R_j + v_i + \epsilon_{ij})},$$

where $y_{ij}$ is a binary indicator of whether the $i$-th group on the $j$-th task improved its collective estimate after social interaction, $\beta_0$ is the fixed intercept, $\beta_1$ is the fixed coefficient for our proposed feature of the estimation context, $v_i$ is the random coefficient for the $i$-th group, and $\epsilon_{ij}$ is a Gaussian error term.

The regression equation for Figure 3.D (Table S1, Model 2) is:

$$y_{ij} = \beta_0 + \beta_1 R_j + \beta_2 I_i + \beta_3 I_i R_j + v_i + \epsilon_{ij},$$

where $y_{ij}$ is the standardized (z-score) absolute error of the revised collective estimate for the $i$-th group in the $j$-th estimation context, $\beta_0$ is the fixed intercept for the regression model, $\beta_1$ is the fixed coefficient for our proposed feature of the estimation context, $I_i \in \{0, 1\}$ is an indicator variable of whether social interaction has occurred or not, $\beta_2$ is the fixed coefficient for the social influence centralization, $\beta_3$ is the fixed coefficient for the interaction term between our proposed feature of the estimation context and influence centralization, and $v_i$ is the random coefficient for the $i$-th group. Finally, $\epsilon_{ij}$ is a Gaussian error term.

Table S1: The main effects of task environment (relative log likelihood, $R$) and the interaction with social influence (i.e., centralization) Each datapoint is an experimental trial. The results are from a mixed effect model with a random effect for the group. Tables S3-S2 for robustness to these choices.

|  | *Dependent variable:* | |
|---|---|---|
|  | Group improved after social interaction | Standardized Absolute Error |
|  | *generalized linear mixed-effects* | *linear mixed-effects* |
|  | (Model 1) | (Model 2) |
| $R$ | 5.543*** | 4.258*** |
|  | (1.053) | (1.171) |
| Social Influence $\in \{0,1\}$ |  | 2.522*** |
|  |  | (0.738) |
| $R$ x Social Influence |  | −4.969*** |
|  |  | (1.257) |
| Intercept | −2.973*** | −2.175*** |
|  | (0.601) | (0.688) |
| Observations | 582 | 687 |
| Log Likelihood | −386.148 | −934.360 |
| Akaike Inf. Crit. | 778.295 | 1,880.720 |
| Bayesian Inf. Crit. | 791.395 | 1,907.914 |

*Note:* *p<0.1; **p<0.05; ***p<0.01

# S4 Robustness checks

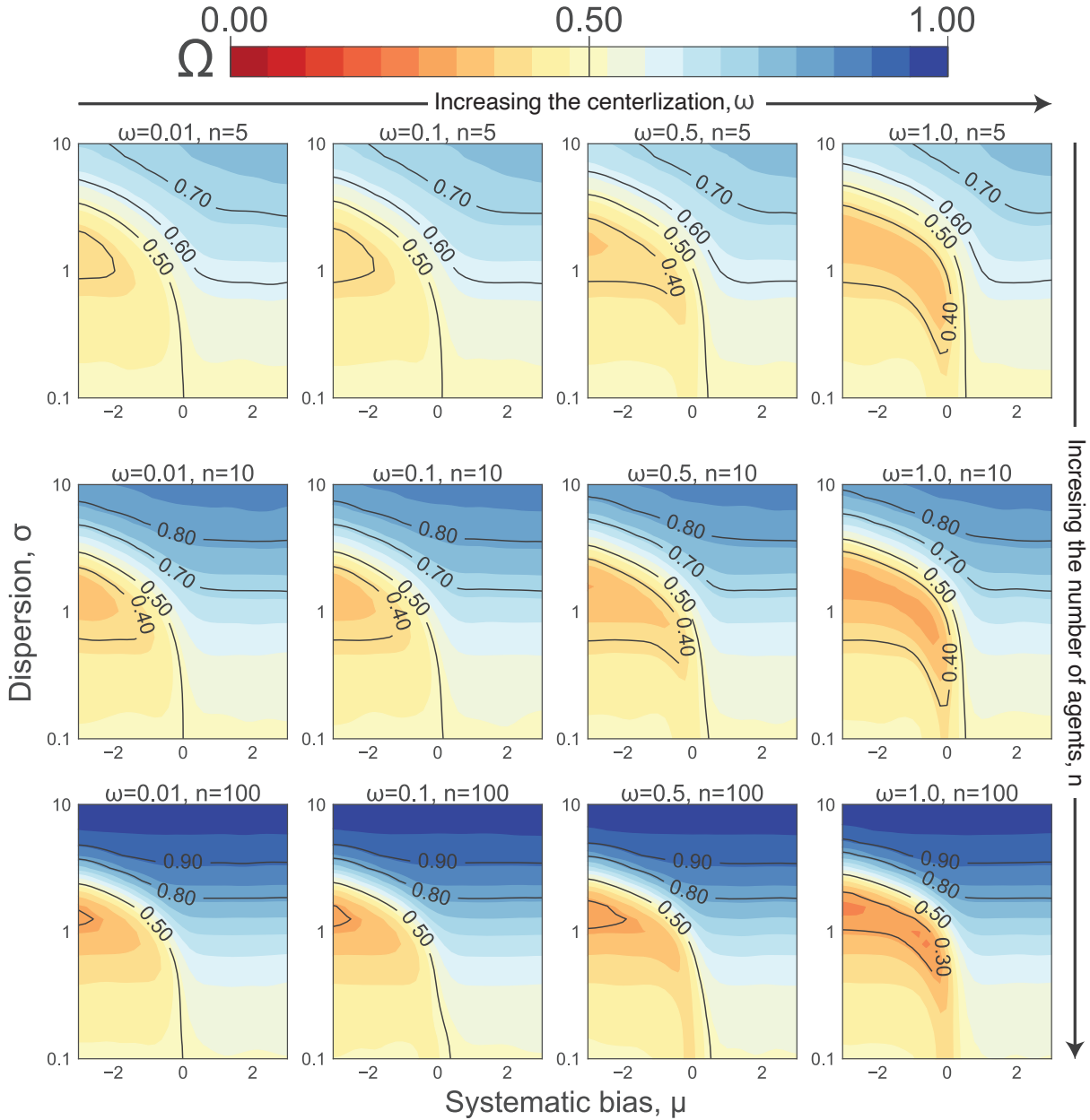

Figure S3: Robustness checks of the simulation results by varying $\omega$ and $n$ when calculating our proposed feature of the estimation context, $\Omega$, for the log-normal distribution. We find that the qualitative behavior of the phase diagram is robust to these changes. Increasing $n$ or $\omega$ leads to sharper transitions from low $\Omega$ to high $\Omega$.

Table S2: Robustness checks for Model 1 (by varying the task environment metric, the independent variable *IV*, derived from the empirical data) for the effect of the task environment on group performance after social interaction. Each datapoint is an experimental trial. Note that for the Hill estimate of the tail index, the average estimate across all possible thresholds was used as the IV, and lower estimates indicate heavier tails. The results are from a mixed effect model with a random effect for the group. We find that the nature of the results is robust across alternatives metrics.

|  | *Dependent variable:* | | |
| --- | --- | --- | --- |
|  | Whether the group improved after social interaction | | |
|  | *Kurtosis* | *Skewness* | *Hill Estimate* |
|  | (1) | (2) | (3) |
| IV | 0.001*** | 0.037*** | −0.156*** |
|  | (0.0003) | (0.009) | (0.044) |
|  |  |  |  |
| Intercept | 0.0004 | −0.150 | 0.621*** |
|  | (0.097) | (0.113) | (0.162) |
| Observations | 582 | 582 | 582 |
| Log Likelihood | −394.647 | −391.573 | −394.093 |
| Akaike Inf. Crit. | 795.294 | 789.147 | 794.186 |
| Bayesian Inf. Crit. | 808.393 | 802.246 | 807.285 |
| *Note:* | | | *p<0.1; **p<0.05; ***p<0.01 |

Table S3: Robustness checks for Model 1 (by varying the task environment metric, the independent variable *IV*, derived from the empirical data) for the marginal effect of the interaction term between the centralization of influence and task environment on group performance—in terms of standardized absolute error. Each datapoint is an experimental trial. Note that for the Hill estimate of the tail index, the average estimate across all possible thresholds was used as the IV, and lower estimates indicate heavier tails. The results are from a mixed effect model with a random effect for the group.

|  | *Dependent variable:* | | |
|---|---|---|---|
|  | Standardized Absolute Error | | |
|  | *Kurtosis* | *Skewness* | *Hill Estimate* |
|  | (1) | (2) | (3) |
| IV | −0.0001 | 0.005 | −0.219*** |
|  | (0.0002) | (0.008) | (0.052) |
| Social Influence ∈ {0,1} | −0.393*** | −0.279* | −0.938*** |
|  | (0.125) | (0.143) | (0.165) |
| IV x Social Influence | 0.0001 | −0.008 | 0.249*** |
|  | (0.0003) | (0.009) | (0.055) |
| Constant | 0.336*** | 0.242* | 0.800*** |
|  | (0.115) | (0.132) | (0.153) |
| Observations | 687 | 687 | 687 |
| Log Likelihood | −958.715 | −951.426 | −938.305 |
| Akaike Inf. Crit. | 1,929.430 | 1,914.851 | 1,888.610 |
| Bayesian Inf. Crit. | 1,956.624 | 1,942.045 | 1,915.804 |

*Note:* *p<0.1; **p<0.05; ***p<0.01

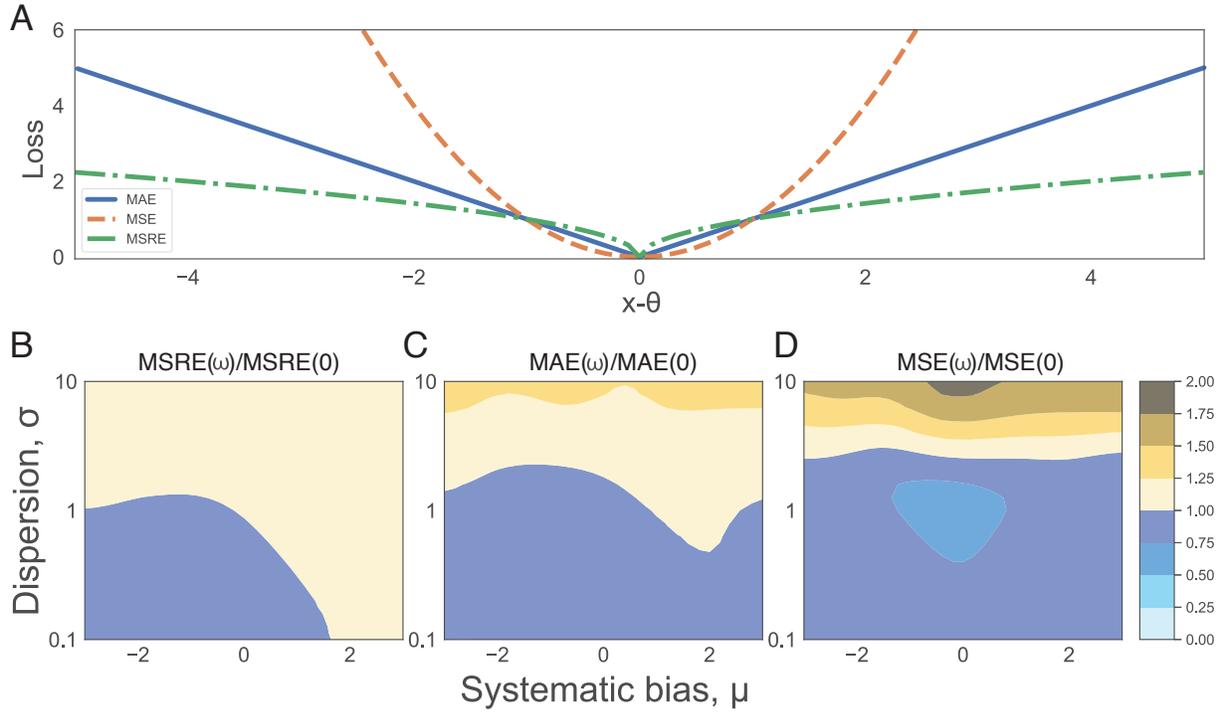

Figure S4: Panel **A** shows the loss as a function of error for the Mean Squared Error (MSE), Mean Absolute Error (MAE), and the Mean Square Root Error (MSRE) loss functions — see (S.6). Panels **B**, **C**, and **D** show the three loss functions for different values of $\mu$ and $\sigma$ for a log-normal distribution. In each case, we plot the ratio of the loss function evaluated in a centralized structure, $\omega > 0$, over a decentralized structure $\omega = 0$. A ratio less than 1 indicates that the centralized network performs better than the decentralized network. The performance of the two influence structures can vary significantly as a function of the selected loss function. The choice of the loss function is typically application-dependent. For instance, if the reward for 'getting it right' is greater than the cost of being frequently wrong — as in domains where the loss and payoff are asymmetric, unbounded, or have a remote boundary [13, 19] — then the decentralized influence structure is more desirable when the dispersion is high. The initial estimates in these simulations are sampled from a log-normal distribution for a fixed number of agents ($n = 50$) and centralization level ($\omega = 1/3$).



\end{document}